\pdfoutput=1

\documentclass[11pt]{article}
\usepackage{array}
\usepackage[final]{acl}

\usepackage{times}
\usepackage{latexsym}
\usepackage{amsmath} 
\usepackage[T1]{fontenc}

\usepackage[utf8]{inputenc}
\usepackage{siunitx}
\usepackage{microtype}
\usepackage{subcaption}
\usepackage{inconsolata}
\usepackage{multirow}
\usepackage{graphicx}
\usepackage{booktabs}
\title{CLaMP 3: Universal Music Information Retrieval Across \\ Unaligned Modalities and Unseen Languages}

\author{
    {\makebox[0.18\textwidth]{\textbf{Shangda Wu}} \hfill
    \makebox[0.18\textwidth]{\textbf{Zhancheng Guo}} \hfill
    \makebox[0.18\textwidth]{\textbf{Ruibin Yuan}} \hfill
    \makebox[0.18\textwidth]{\textbf{Junyan Jiang}} \hfill
    \makebox[0.18\textwidth]{\textbf{Seungheon Doh}}} \hfill
    \\[0.5ex]
    \makebox[0.18\textwidth]{\textbf{Gus Xia}} \hfill
    \makebox[0.18\textwidth]{\textbf{Juhan Nam}} \hfill
    \makebox[0.18\textwidth]{\textbf{Xiaobing Li}} \hfill
    \makebox[0.18\textwidth]{\textbf{Feng Yu}} \hfill
    \makebox[0.18\textwidth]{\textbf{Maosong Sun}}
    \\[1ex]
    {\small Details of authors, correspondence, and affiliations are on Page 9}
    \\[1.5ex]
    \url{https://sanderwood.github.io/clamp3}}

\begin{document}
\maketitle
\begin{abstract}
CLaMP 3 is a unified framework developed to address challenges of cross-modal and cross-lingual generalization in music information retrieval. Using contrastive learning, it aligns all major music modalities--including sheet music, performance signals, and audio recordings--with multilingual text in a shared representation space, enabling retrieval across unaligned modalities with text as a bridge. It features a multilingual text encoder adaptable to unseen languages, exhibiting strong cross-lingual generalization. Leveraging retrieval-augmented generation, we curated M4-RAG, a web-scale dataset consisting of 2.31 million music-text pairs. This dataset is enriched with detailed metadata that represents a wide array of global musical traditions. To advance future research, we release WikiMT-X, a benchmark comprising 1,000 triplets of sheet music, audio, and richly varied text descriptions. Experiments show that CLaMP 3 achieves state-of-the-art performance on multiple MIR tasks, significantly surpassing previous strong baselines and demonstrating excellent generalization in multimodal and multilingual music contexts.
\end{abstract}

\section{Introduction}
Music Information Retrieval (MIR) is a field that aims at developing computational tools for processing, organizing, and accessing music data. A core challenge in MIR is retrieving musical content—whether sheet music, performance signals, or audio recordings—based on natural language queries (“a fast-paced classical piano piece”). This connection enables applications such as automatic music tagging, where models assign genres (“jazz,” “folk”) or descriptive attributes (“melancholic,” “upbeat”), facilitating music organization, search, and recommendation. By integrating NLP methodologies, MIR enables more intuitive access to musical content, making it more interpretable and searchable through text.

\begin{figure}[t]
    \centering
    \includegraphics[width=0.475\textwidth]{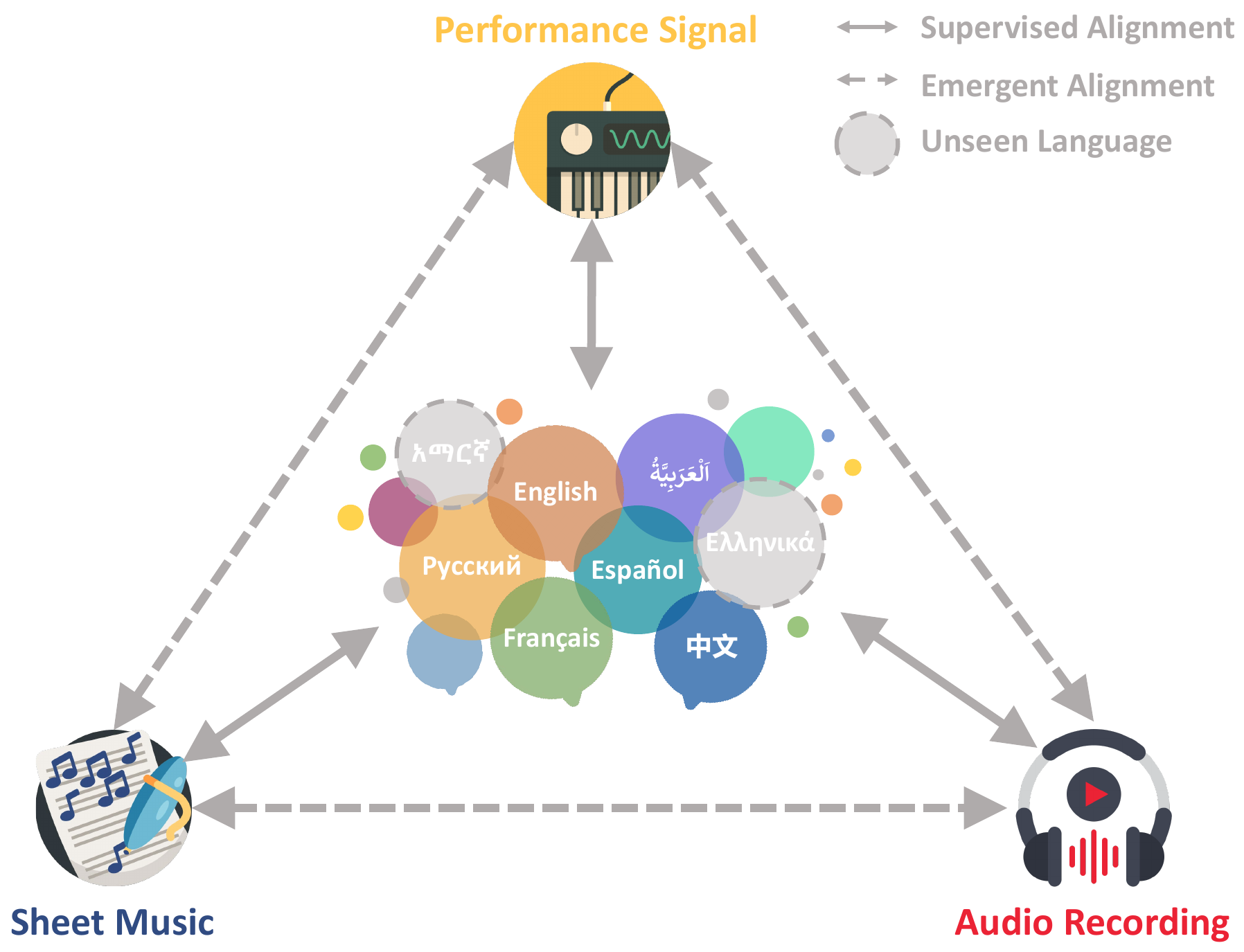}
    \caption{CLaMP 3 demonstrates robust cross-modal and cross-lingual generalization. Supervised alignment (solid arrows) links paired modalities, while emergent alignment (dashed arrows) bridges unaligned ones. A multilingual text encoder enables retrieval in languages unseen (grayed-out bubbles) during alignment.}
    \vspace{-1em}
    \label{fig:overview}
\end{figure}

These capabilities position MIR as a critical bridge between music and language, supporting various applications beyond retrieval and annotation. For instance, cross-modal representations enable text-to-music generation models \cite{agostinelli2023musiclm, DBLP:conf/icassp/ChenWLNBD24} to create music based on text descriptions. MIR also aids in the automatic evaluation of these models by assessing how closely the generated music aligns with text descriptions or resembles the ground truth \cite{DBLP:conf/nips/CopetKGRKSAD23, retkowski2024frechet}.

Despite these advancements, MIR faces significant challenges in addressing the complexities of \textbf{\textit{multimodality}} and \textbf{\textit{multilinguality}}. Music exists in many forms: sheet music offers human-readable representations for theoretical analysis and education; performance signals (e.g., MIDI) capture timing and dynamics for precise digital editing; and audio recordings serve as the primary medium for listening. While these modalities complement each other, their heterogeneous representational structures complicate unified computational processing.

Adding to this complexity, as a universal medium, music is described in numerous languages, crossing cultural and linguistic boundaries. Musical terminology, descriptions, and cultural references vary significantly between linguistic communities, each bringing its own rich vocabulary and cultural context. To build global and accessible MIR systems, it is essential to process and understand these diverse expressions effectively.

\begin{figure*}[t]
    \centering
    \includegraphics[width=\textwidth]{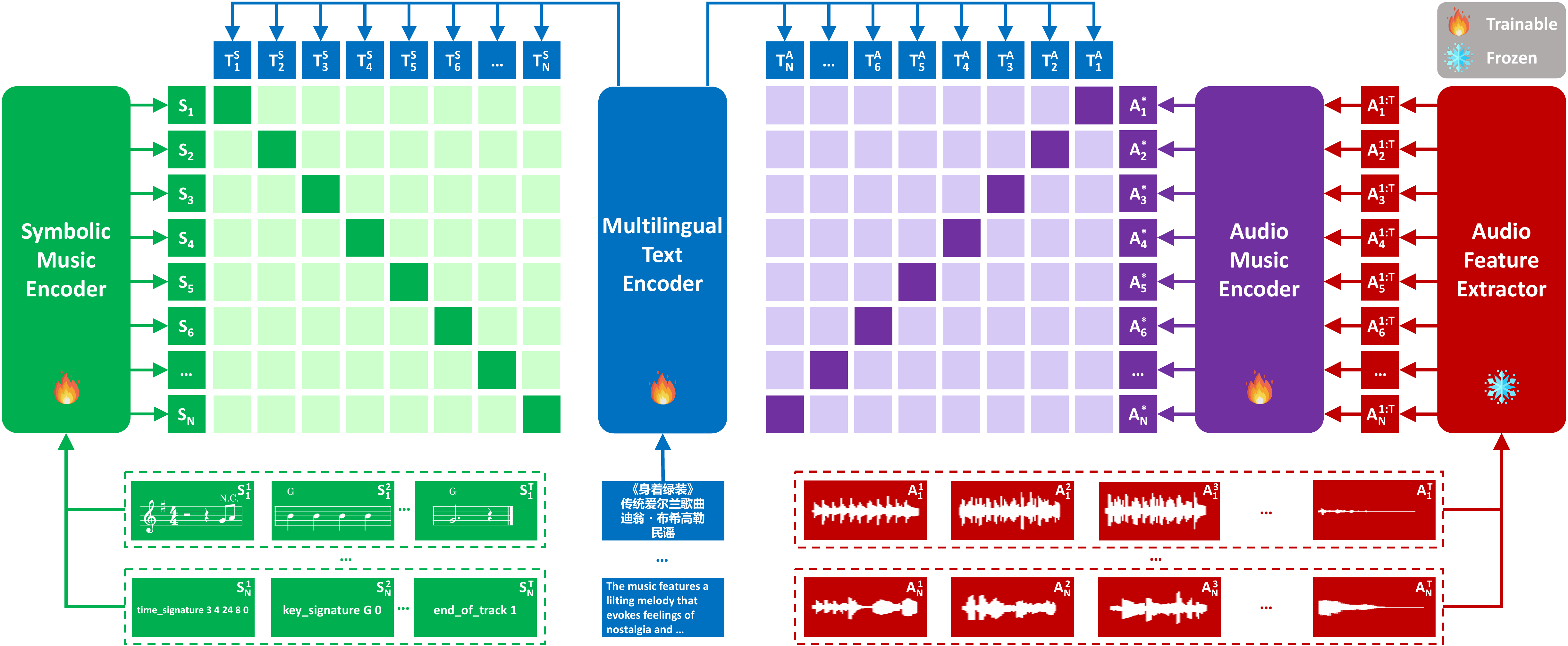}
    \caption{CLaMP 3 uses contrastive learning to align features across modalities. Sheet music and performance signals are segmented into units (bars or MIDI messages) and processed by the symbolic music encoder, while audio is segmented into 5-second clips and processed through the audio feature extractor and audio music encoder. Both symbolic and audio representations are aligned with text representations from the multilingual text encoder.}
    \vspace{-1em}
    \label{fig:clamp3}
\end{figure*}

Unfortunately, the development of MIR is limited not only by the lack of music-text pairs but also by the general scarcity of paired data across different musical modalities. As a result, most research focuses on retrieval between specific modality pairs, such as text and audio \cite{DBLP:conf/ismir/HuangJLGLE22,DBLP:conf/icassp/DohLJN24,zhu2025muq} or text and sheet music \cite{DBLP:conf/ismir/WuY0S23}. This narrow focus restricts the potential for cross-modal interactions, preventing a more comprehensive understanding of music. Additionally, existing text data is often short-form, like tags, with few long-form descriptions \cite{DBLP:conf/icassp/WuCZHBD23}, leading to shallow semantics. These datasets are also predominantly in English \cite{DBLP:conf/icassp/DohWCN23a}, with limited representation of other languages, neglecting music's global and multilingual nature.

To tackle these challenges, a unified framework is crucial for aligning musical modalities and bridging linguistic gaps, particularly in the absence of paired training data. Large Language Models (LLMs) present a promising solution by addressing the limitations of text semantics and the scarcity of linguistic diversity in music-text datasets. These models excel at transforming basic metadata into fluent and contextually rich descriptions \cite{DBLP:conf/ismir/DohCLN23,bai2024audiosetcaps}. Furthermore, their multilingual capabilities allow them to support a wide array of languages \cite{wu2024clamp}, enhancing semantic depth and enabling more inclusive access across diverse linguistic and cultural contexts.

In this paper, we introduce CLaMP 3, a universal MIR framework that processes music and text while aligning them into a shared representation space. It covers all major music modalities: 1) sheet music, 2) performance signals, and 3) audio recordings, along with 4) multilingual text. Each modality is encoded through its respective feature extractor. To unify these representations, we employ contrastive learning \cite{DBLP:conf/nips/Sohn16}, aligning both musical and textual features. This enables seamless cross-modal retrieval and integration across diverse musical formats and languages.

To address the shortage of paired music-text data, we use Retrieval-Augmented Generation (RAG) \cite{DBLP:conf/nips/LewisPPPKGKLYR020} to create M4-RAG, a dataset of 2.31 million music-text pairs covering various musical modalities. Starting with basic metadata like song titles and artist names, we retrieve relevant web documents and use an LLM to generate detailed annotations. These annotations include short tags, long descriptions, and multilingual translations, providing rich and diverse information.

In addition, we present WikiMT-X, the first benchmark to align text, audio, and sheet music. It includes 1,000 triplets with diverse text annotations, such as genre labels and detailed long-form descriptions, including background context, musical analysis, general descriptions, and scene depictions. WikiMT-X facilitates evaluation across modalities and semantic perspectives, providing a holistic framework to assess models’ ability to align and interpret musical content.

Experiments demonstrate that CLaMP 3 achieves state-of-the-art performance on various MIR tasks, including text-to-audio and text-to-symbolic music retrieval, significantly surpassing all baselines. It also excels in multilingual retrieval, generalizing to languages not present during alignment. By leveraging text as a bridge, CLaMP 3 enables emergent cross-modal retrieval, connecting musical modalities without paired training data.

Overall, this work contributes:

\begin{itemize}
    \item CLaMP 3 unifies musical modalities and languages in a shared representation space, achieving strong performance on a wide range of MIR tasks and generalizing to unseen languages with emergent cross-modal alignment.

    \item We curate M4-RAG, a dataset of 2.31 million music-text pairs with diverse annotations, spanning 27 languages and 194 countries, addressing a critical gap in high-quality training data for music and language tasks.

    \item WikiMT-X links text, audio, and sheet music with 1,000 triplets, offering a first-of-its-kind resource to evaluate models holistically across different modalities and semantic aspects.
    
\end{itemize}

To support future research, we have publicly released the complete codebase, pre-trained weights of CLaMP 3, 1.56 million audio-text training pairs, and the WikiMT-X benchmark\footnote{\url{https://github.com/sanderwood/clamp3}}.

\section{Model}
\subsection{Training Objective}
\label{sec:training}
CLaMP 3’s optimization objective is to minimize the InfoNCE loss \cite{oord2018representation}, aligning embeddings using contrastive learning:

\begingroup
\fontsize{10pt}{9pt}\selectfont
\begin{equation}
\begin{split}
L_{CL} = - \frac{1}{N} \sum_{i=1}^{N} \log \frac{\exp(\text{sim}(z_i^{t}, z_i^{m}) / \tau)}{\sum_{j=1}^{N} \exp(\text{sim}(z_i^{t}, z_j^{m}) / \tau)},
\end{split}
\end{equation}
\endgroup

\noindent
where \( z_i^{t} \) and \( z_i^{m} \) are text and music embeddings, \(\text{sim}(\cdot, \cdot)\) is the similarity function (e.g., dot product), and \(\tau\) is the temperature parameter. Positive pairs are aligned text-music samples, while negatives are unrelated samples from the same batch.

Inspired by ImageBind \cite{DBLP:conf/cvpr/GirdharELSAJM23}, we adopt a multi-stage strategy using text as a bridge to address the lack of paired music data:

\textbf{Stage 1:} The text encoder is first trained to align with one music encoder (e.g., symbolic encoder).

\textbf{Stage 2:} It is then aligned with another music encoder (e.g., audio encoder), freezing the text encoder to prevent representation drift.

\textbf{Stage 3:} The text encoder is unfrozen to refine its alignment with the music encoder from Stage 2.

\textbf{Stage 4:} The text encoder is frozen again to prevent shifts while re-aligning with the Stage 1 music encoder to fix alignment drift from Stage 3.

This strategy minimizes modality interference while mapping all modalities into a shared representation space for effective cross-modal transfer.

\subsection{Core Components}
CLaMP 3 consists of several transformer-based encoders \cite{DBLP:conf/nips/VaswaniSPUJGKP17} for each modality:

\textbf{Multilingual Text Encoder:} The text encoder in CLaMP 3 is based on XLM-R-base \cite{DBLP:conf/acl/ConneauKGCWGGOZ20}, a model pre-trained on 2.5 TB of CommonCrawl data across 100 languages. It has 12 layers and a hidden size of 768, enabling strong cross-lingual generalization to unseen languages.

\textbf{Symbolic Music Encoder:} CLaMP 3 uses M3 \cite{wu2024clamp}, a self-supervised model for encoding symbolic music, including multi-track voice-interleaved ABC notation and lossless MIDI encoding via MIDI Text Format (MTF). M3 segments ABC into bars and MIDI into messages, treating each segment as a patch. The model has 12 encoder layers, a hidden size of 768, and processes up to 512 patches or 32,768 characters per input.

\textbf{Audio Music Encoder:} It is a 12-layer transformer with a 768-dimensional hidden size, trained from scratch for audio processing. This encoder leverages pre-trained features from MERT-v1-95M \cite{DBLP:conf/iclr/LiYZMCYXLRBGDLC24}, where MERT serves as a frozen audio feature extractor. Each 5-second clip is represented by a single embedding, obtained by averaging across all MERT layers and time steps. CLaMP 3 processes up to 128 such embeddings, covering 640 seconds of audio, allowing it to capture high-level audio patterns over extended durations.

All encoders process their outputs through a linear layer, followed by average pooling, to generate a single global semantic feature for each input.

\section{Dataset}
In this section, we introduce the M4-RAG dataset for training CLaMP 3 and the WikiMT-X benchmark for evaluation. We start with data sources, followed by the metadata curation process. Then, we summarize dataset statistics like scale and diversity. Finally, we elaborate on the details of the WikiMT-X benchmark.

\subsection{Data Sources}
The training data for CLaMP 3 is built from both symbolic and audio music datasets, ensuring a rich and diverse foundation for multimodal learning.

The symbolic music data is sourced from WebMusicText (WebMT) \cite{DBLP:conf/ismir/WuY0S23} with 1.4 million ABC notation files and the Million MIDI Dataset (MMD) \cite{DBLP:conf/acl/ZengTWJQL21} with 1.5 million MIDI files. Since symbolic music formats use discrete symbols to represent music, they can be converted into one another, albeit with some information loss. To fully utilize the data, these datasets were unified by converting MMD to ABC and WebMT to MIDI. This process yields 3 million symbolic music files, offering diverse and comprehensive training coverage.

The audio data is collected from online sources, comprising 160 thousand hours of audio from 1.8 million tracks. As CLaMP 3 directly utilizes pre-extracted features, the training data exclusively consists of these precomputed features, leading to substantial savings in both computational resources and time.

\begin{table}[t]
\centering
\renewcommand{\arraystretch}{1.25}
\fontsize{7.85}{8}\selectfont
\setlength{\tabcolsep}{4pt}
\caption{Metadata overview for M4-RAG, grouped into basic information, annotations, and translations. In \textbf{\textit{Annotations}}, \textit{Region} and \textit{Language} are written in English; other fields follow the \textit{Language} specification.}
\begin{tabular}{l l l r}
\toprule
\textbf{Category}       & \textbf{Field}         & \textbf{Content}             & \textbf{Avg Bytes} \\
\midrule
\multirow{2}{*}{\textbf{\textit{Basic}}}          
                        & \textit{Title}        & Music Title                   & 20.04 \\
                        & \textit{Artists}      & Artist names                  & 21.97 \\
\midrule
\multirow{6}{*}{\textbf{\vspace{-8mm}\textit{Annotations}}}    
                        & \textit{Region}       & Country of origin             & 20.69 \\
                        & \textit{Language}     & Document language              & 7.02 \\
                        & \textit{Genres}       & Genre list                    & 21.83 \\
                        & \textit{Tags}         & Keywords/playlists            & 51.91 \\
                        & \textit{Background}   & Background context            & 531.79 \\
                        & \textit{Analysis}     & Musical analysis              & 770.29 \\
                        & \textit{Description}  & General description           & 591.86 \\
                        & \textit{Scene}        & Scene depiction               & 750.92 \\
\midrule
\multirow{5}{*}{\textbf{\textit{Translations}}}   
                        & \textit{Language}     & Translation language          & 6.38 \\
                        & \textit{Background}   & Translated background         & 819.76 \\
                        & \textit{Analysis}     & Translated analysis           & 1130.47 \\
                        & \textit{Description}  & Translated description        & 888.86 \\
                        & \textit{Scene}        & Translated scene              & 1077.07 \\
\bottomrule
\end{tabular}
\label{tab:dataset_fields}
\end{table}

\subsection{Metadata Curation}
Music titles often serve as unique identifiers, enabling the retrieval of rich and detailed descriptions from diverse online sources. When paired with artist names, they further refine searches, pinpointing specific versions or performances and reducing ambiguities caused by covers or adaptations. This distinctive property makes music titles a reliable basis for generating annotations, even in the absence of paired music-text datasets.

To leverage this, we curated M4-RAG (Million-scale Multilingual Music Metadata), a dataset comprising 2.31 million metadata entries. The curation process involved several key steps:

\textbf{Title Filtering:} Entries without titles were excluded, as titles are essential for retrieving meaningful information from the web.

\textbf{Web Search:} Google searches were conducted using titles and, where available, artist names. For each entry, the top 10 search results were collected to ensure diverse and reliable sources.

\textbf{RAG:} Using Qwen2.5-72B \cite{yang2024qwen2}, we generated annotations from the retrieved documents and basic metadata (titles and artist names). The annotations covered the fields in Table~\ref{tab:dataset_fields} under \textbf{\textit{Annotations}}, with an additional Boolean field indicating if the source material had sufficient information for generating meaningful annotations.

\textbf{Quality Filtering:} Entries were discarded if flagged by the Boolean field for insufficient information, if their format failed to meet the standards outlined in Table~\ref{tab:dataset_fields}, or if any fields were left empty.

\textbf{Postprocessing:} To address inconsistencies in the generated annotations, \textit{Region} fields were mapped to recognized countries, while \textit{Description} fields were refined using Qwen to remove identifiable details such as titles and lyrics. Language consistency across long-form fields (\textit{Background}, \textit{Analysis}, \textit{Description}, \textit{Scene}) was verified with fastText \cite{DBLP:conf/eacl/GraveMJB17}. Entries with inconsistent languages or languages unsupported by either XLM-R or Qwen were removed, and valid detected languages were recorded in the \textit{Language} field.

\textbf{Multilingual Translation:} To enhance linguistic diversity, a random language supported by both XLM-R and Qwen—different from the original—was selected for each entry, and long-form annotations were translated into it using Qwen.

Prompt and examples of generated annotations are provided in Appendix~\ref{sec:prompt}.

\begin{figure}[t]
    \centering
    \includegraphics[width=0.475\textwidth]{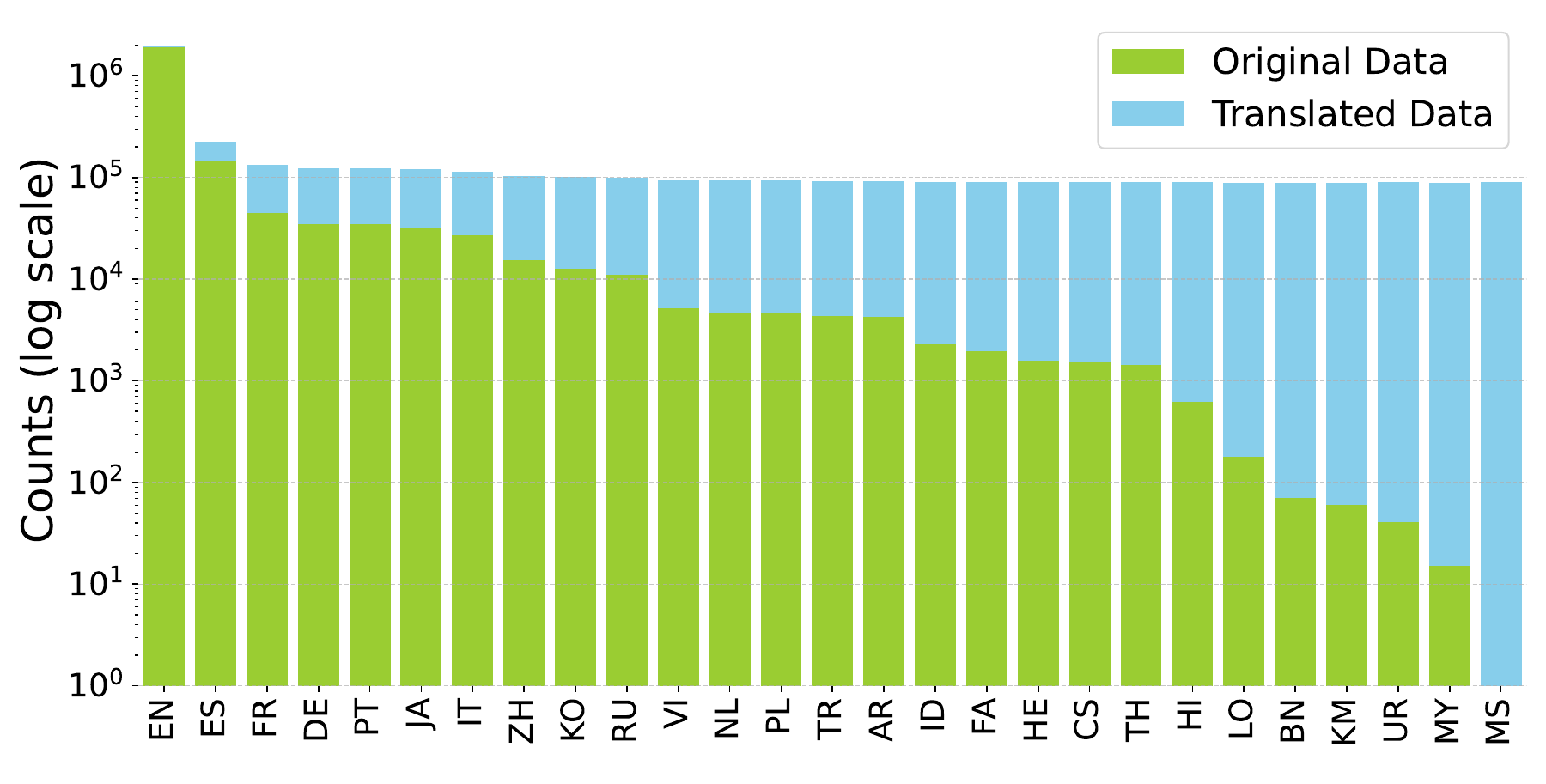}
    \caption{Language distribution of original and translated entries in M4-RAG, covering 27 languages.}
    \label{fig:languages}
\end{figure}

\begin{figure}[t]
    \centering
    \includegraphics[width=0.475\textwidth]{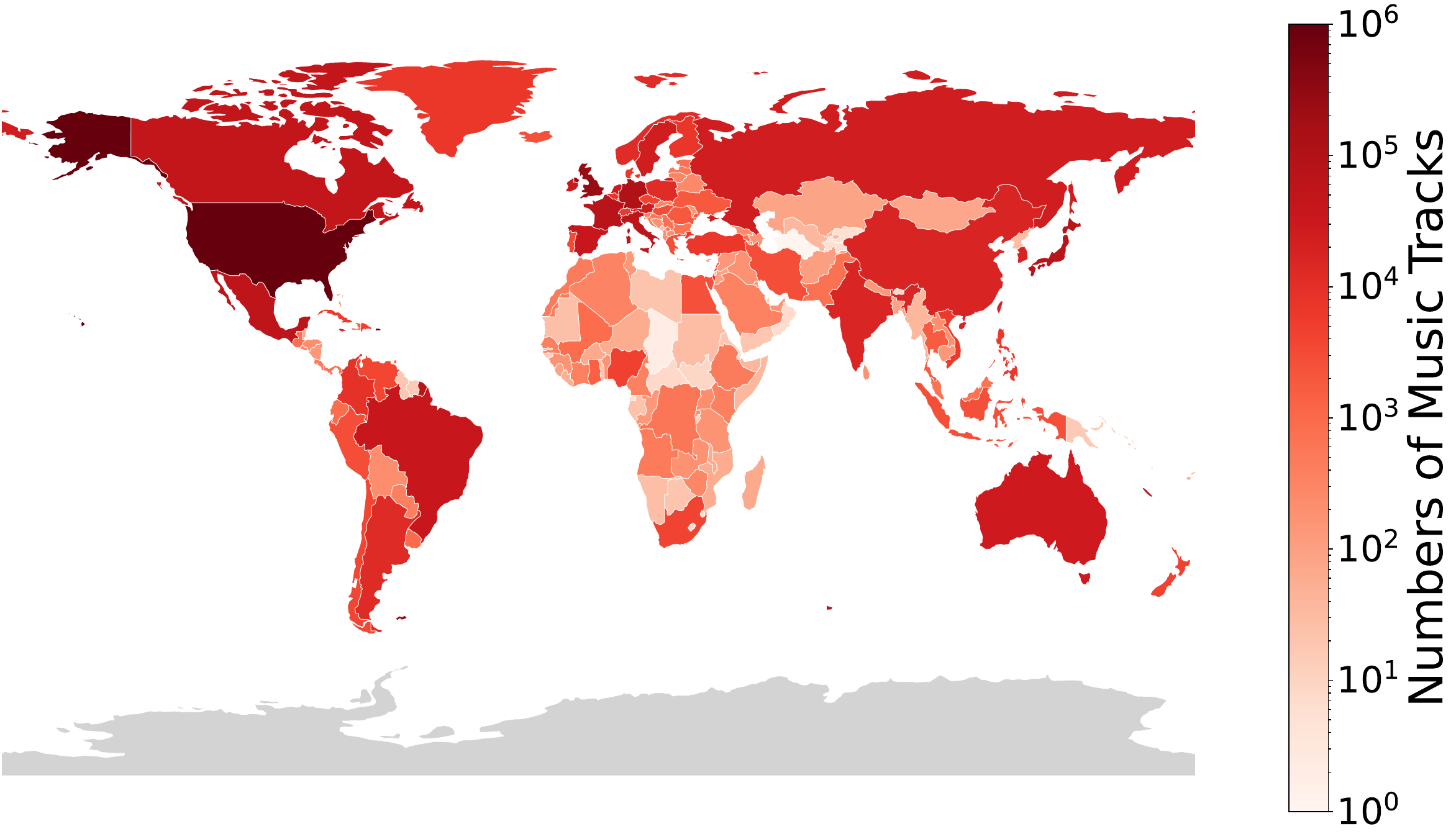}
    \caption{Country-wise distribution of music tracks in M4-RAG, spanning 194 countries.}
    \label{fig:regions}
\end{figure}

\subsection{Dataset Statistics}
Through metadata curation, we obtained M4-RAG, which consists of 2.31 million entries. It includes 0.58 million ABC-text pairs from WebMT, 0.17 million MIDI-text pairs from MMD, and 1.56 million audio-text pairs.

Each metadata entry includes both short-form annotations, such as genres and tags, and detailed long-form descriptions. As summarized in Table~\ref{tab:dataset_fields}, the long-form descriptions account for the majority of the dataset, providing extensive semantic details from multiple perspectives.

M4-RAG spans 27 languages, with the original metadata predominantly in English, as shown in Fig.~\ref{fig:languages}. To address this imbalance, translations were added to the long-form descriptions, greatly boosting non-English data. This was particularly impactful for low-resource languages, such as Malay and Burmese, where most data depends on translations, greatly enhancing their representation.

In terms of geographic coverage, M4-RAG incorporates music from 194 countries. Fig.~\ref{fig:regions} illustrates contributions from both major music-producing nations and less-represented regions. This global reach ensures the dataset reflects a diverse range of musical traditions and styles from across the world.

\begin{figure}[t]
    \centering
    \includegraphics[width=0.475\textwidth]{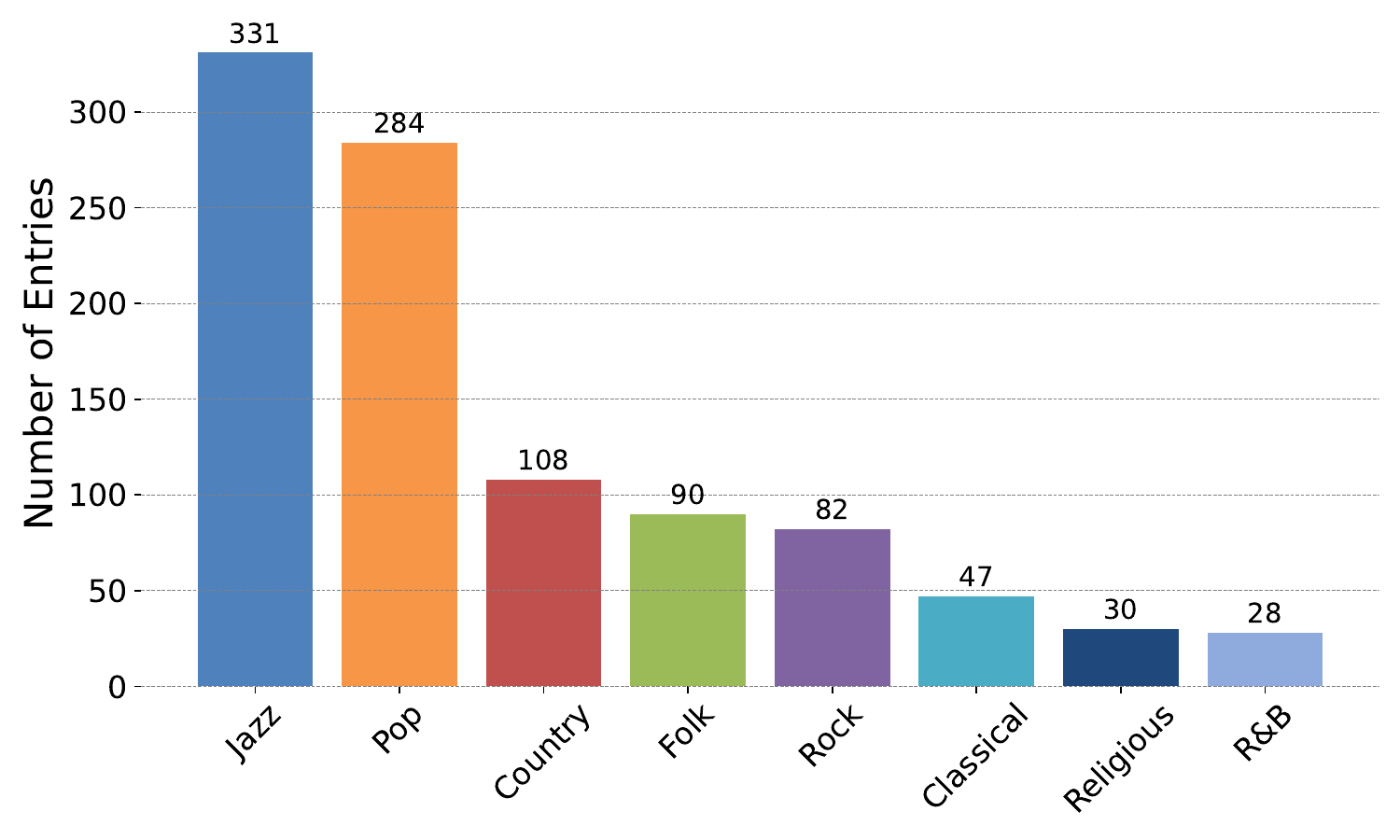}
    \vspace{-0.8em}
    \caption{Genre distribution of the WikiMT-X dataset.}
    \vspace{-1em}
\end{figure}

\subsection{Benchmark Dataset}
WikiMT-X (WikiMusicText-eXtended) extends WikiMT \cite{DBLP:conf/ismir/WuY0S23}, focusing on 20th-century Western music with 1,000 entries, each with sheet music, audio, and detailed metadata.

The original WikiMT dataset had the following drawbacks: 1) the text was sourced from Wikipedia, mainly focused on background information with limited semantic diversity; 2) the absence of audio data severely restricted the evaluation scope; and 3) the genre labels were obtained through keyword matching, resulting in relatively low accuracy and reducing the reliability of the dataset.

To address these deficiencies, WikiMT-X made the following improvements:

\begin{itemize}
    \item We used llama-3.1-sonar-large-128k-online\footnote{\url{https://www.perplexity.ai}} \cite{dubey2024llama}, feeding it sheet music with titles, artist names, and lyrics. It retrieved relevant web pages and summarized them into background, analysis, description, and scene.
    \item We manually matched sheet music with audio recordings retrieved from YouTube and removed 10 identified duplicates.
    \item We reorganized genre categories based on data distribution and re-annotated labels.
\end{itemize}

These enhancements make WikiMT-X useful for multimodal MIR research tasks, assessing models’ capabilities in handling text annotations of diverse semantic types, and classifying music across modalities using genre labels.

Appendix~\ref{sec:eval_benchmark} presents detailed objective and human evaluations of WikiMT-X annotation quality. In addition, Appendix~\ref{sec:t-sne} provides \textit{t}-SNE visualizations of CLaMP 3 embeddings on WikiMT-X, showing modality, language, and semantic distributions in the shared representation space.

\begin{table*}[t]
\centering
\renewcommand{\arraystretch}{1.25}
\setlength{\tabcolsep}{12.5pt} 
\fontsize{9.7}{8}\selectfont
\caption{Results for English text-to-music retrieval on several benchmarks: WikiMT and MidiCaps have 1,010 pairs, Song Describer Dataset (SDD) has 706 audio and 1,106 captions, and MusicCaps-Remake (MC-R) contains 2,777 pairs. MC-R prevents data leakage by using full-length audio and rewritten captions from AudioSet’s evaluation set.}
\begin{tabular}{l c c c c c c}
\toprule
\multirow{2}{*}{\textbf{\vspace{-2mm}Model}} & \multicolumn{2}{c}{\textbf{Symbolic Benchmarks}} & \multicolumn{4}{c}{\textbf{WikiMT-X (Sheet Music)}} \\ 
\cmidrule(lr){2-3} \cmidrule(lr){4-7}
& \textit{WikiMT} & \textit{MidiCaps} & \textit{Background} & \textit{Analysis} & \textit{Description} & \textit{Scene} \\
\midrule
\textit{CLaMP}           & 0.2561 & 0.1236 & 0.2122 & 0.1345 & 0.0306 & 0.0426 \\
\textit{CLaMP 2}         & 0.3438 & 0.2695 & 0.3024 & 0.2374 & 0.0418 & 0.0838 \\
\textit{CLaMP 3$_{sa}^{c2}$}  & \textbf{0.4498} & \textbf{0.2826} & \textbf{0.4028} & \textbf{0.3382} & 0.0835 & \textbf{0.1512} \\
\textit{CLaMP 3$_{saas}$}  & 0.3555 & 0.1798 & 0.3301 & 0.2758 & \textbf{0.1274} & 0.1500 \\
\midrule
\midrule
\multirow{2}{*}{\textbf{\vspace{-2mm}Model}} & \multicolumn{2}{c}{\textbf{Audio Benchmarks}} & \multicolumn{4}{c}{\textbf{WikiMT-X (Audio)}} \\ 
\cmidrule(lr){2-3} \cmidrule(lr){4-7}
& \textit{SDD} & \textit{MC-R} & \textit{Background} & \textit{Analysis} & \textit{Description} & \textit{Scene} \\
\midrule
\textit{CLAP}            & 0.1310 & 0.0657 & 0.0598 & 0.0429 & 0.0318 & 0.0218 \\
\textit{TTMR++}          & 0.1437 & \textbf{0.1248} & 0.1119 & 0.0833 & 0.0584 & 0.0301 \\
\textit{CLaMP 3$_{sa}^{c2}$}  & 0.1612 & 0.0959 & 0.1180 & 0.1206 & 0.0639 & 0.0619 \\
\textit{CLaMP 3$_{saas}$}  & \textbf{0.1985} & 0.1177 & \textbf{0.2017} & \textbf{0.1711} & \textbf{0.0988} & \textbf{0.0963} \\
\bottomrule
\end{tabular}
\label{tab:model_comparison}
\end{table*}

\section{Experiments}
\label{sec:exp}
This section evaluates CLaMP 3 on retrieval tasks, comparing it to state-of-the-art baselines. We present results for the two best-performing CLaMP 3 variants—one for symbolic music and one for audio. A full retrieval comparison of all variants can be found in Appendix~\ref{sec:variants}, and classification results are available in Appendix~\ref{sec:classification}.

\subsection{Settings}
Both symbolic music and audio alignments were trained for up to 100 epochs on 8 NVIDIA H800 GPUs. Symbolic music alignment required 4 days with a learning rate of 5e-5 and a batch size of 1024. Audio alignment took 1 day with a learning rate of 1e-5 and a batch size of 2048.

M4-RAG was divided into 99\% for training and 1\% for validation. During training, metadata information was randomly selected to form text inputs. Mixed-precision \cite{DBLP:conf/iclr/MicikeviciusNAD18}, AdamW optimizer \cite{DBLP:conf/iclr/LoshchilovH19}, and a 1,000-step warm-up \cite{DBLP:journals/corr/GoyalDGNWKTJH17} were used to enhance efficiency.

Following the training strategy in Sec.~\ref{sec:training}, we explored various modality alignment orders for symbolic and audio modalities, and present the two top-performing variants below:

\textbf{CLaMP 3\(_{\textnormal {saas}}\):} Optimized for audio, this model follows the full multi-stage alignment: symbolic $\rightarrow$ audio $\rightarrow$ audio $\rightarrow$ symbolic.
    
\textbf{CLaMP 3\(_{\textnormal {sa}}^{\textnormal {c2}}\):} Optimized for symbolic, this model starts from CLaMP 2-initialized text and symbolic encoders, followed by two stages: the text encoder is jointly trained with the symbolic encoder, then frozen to align with the audio encoder.

\subsection{English Text-to-Music Retrieval}
We evaluated retrieval performance using Mean Reciprocal Rank (MRR), which measures the inverse of the rank of the paired item, across all tasks.

For symbolic music retrieval, we compared CLaMP 3 with CLaMP 2 \cite{wu2024clamp} and CLaMP \cite{DBLP:conf/ismir/WuY0S23} on WikiMT (using ABC notation) and MidiCaps \cite{melechovsky2024midicaps} (using MIDI). For audio retrieval, we evaluated CLaMP 3 against state-of-the-art models CLAP \cite{DBLP:conf/icassp/WuCZHBD23} and TTMR++ \cite{DBLP:conf/icassp/DohLJN24} on the Song Describer Dataset (SDD) \cite{manco2023song} and MusicCaps-Remake (MC-R) \cite{agostinelli2023musiclm}, which addresses data leakage by using full-length audio and rewritten captions (see Appendix~\ref{sec:musiccaps}) from AudioSet’s evaluation set \cite{DBLP:conf/icassp/GemmekeEFJLMPR17}. In addition, we tested all models on WikiMT-X to evaluate their performance across varying semantic perspectives.

As shown in Table~\ref{tab:model_comparison}, CLaMP 3 achieved significant improvements over its predecessors and baseline models across both symbolic and audio retrieval tasks. For symbolic music retrieval, CLaMP 3\(_{\textnormal {sa}}^{\textnormal {c2}}\) achieved MRR scores of 0.4498 on WikiMT and 0.2826 on MidiCaps, clearly outperforming both CLaMP 2 and CLaMP, despite using only half the training data. This improvement can be attributed to the high-quality, richly annotated M4-RAG dataset. Similarly, CLaMP 3\(_{\textnormal {saas}}\), though optimized for audio retrieval, exceeded CLaMP by a notable margin on symbolic benchmarks and performed comparably to CLaMP 2 on WikiMT. These results demonstrate that our multi-stage training approach effectively preserves performance on modalities that were not explicitly optimized.

\begin{table*}[t]
\centering
\renewcommand{\arraystretch}{1.25}
\setlength{\tabcolsep}{6pt} 
\fontsize{9.7}{8}\selectfont
\caption{Results for multilingual text-to-music retrieval on translated WikiMT-X background annotations. Languages marked with asterisks were not included in the M4-RAG training data. The BLEU scores below each language are calculated by back-translating the text with the SeamlessM4T model and comparing it to the original English text.}
\begin{tabular}{l c c c c c c c c c c}
\toprule
\multirow{2}{*}{\textbf{Model}} & \textbf{ru} & \textbf{fr} & \textbf{es} & \textbf{ar} & \textbf{zh} & \textbf{fi*} & \textbf{el*} & \textbf{ta*} & \textbf{kk*} & \textbf{am*} \\
& \textit{49.69} & \textit{55.50} & \textit{62.82} & \textit{53.38} & \textit{39.58} & \textit{39.19} & \textit{55.55} & \textit{40.07} & \textit{36.57} & \textit{56.08} \\ 
\midrule
\textbf{ABC Notation} & & & & & & & & & & \\
\textit{CLaMP 2}        & 0.2668 & 0.2968 & 0.2934 & 0.2298 & 0.1646 & 0.2795 & 0.2410 & 0.0915 & 0.2543 & 0.1237 \\
\textit{CLaMP 3$_{sa}^{c2}$}  & \textbf{0.3614} & \textbf{0.3949} & \textbf{0.3921} & \textbf{0.3155} & \textbf{0.2373} & \textbf{0.3524} & \textbf{0.3226} & \textbf{0.1415} & \textbf{0.3397} & \textbf{0.1871} \\
\textit{CLaMP 3$_{saas}$}  & 0.2918 & 0.3214 & 0.3239 & 0.2789 & 0.2358 & 0.2919 & 0.2681 & 0.1246 & 0.2703 & 0.1139 \\
\midrule
\textbf{MIDI} & & & & & & & & & & \\
\textit{CLaMP 2}        & 0.1271 & 0.1414 & 0.1452 & 0.1113 & 0.0749 & 0.1438 & 0.1087 & 0.0466 & 0.1079 & 0.0616 \\
\textit{CLaMP 3$_{sa}^{c2}$}  & \textbf{0.1921} & \textbf{0.2101} & \textbf{0.2137} & \textbf{0.1681} & \textbf{0.1316} & \textbf{0.2019} & \textbf{0.1702} & \textbf{0.0804} & \textbf{0.1765} & \textbf{0.1039} \\
\textit{CLaMP 3$_{saas}$}  & 0.1165 & 0.1319 & 0.1330 & 0.1141 & 0.0937 & 0.1245 & 0.1143 & 0.0601 & 0.1104 & 0.0544 \\
\midrule
\textbf{Audio} & & & & & & & & & & \\
\textit{CLaMP 3$_{sa}^{c2}$}  & 0.1068 & 0.1150 & 0.1202 & 0.0981 & 0.0877 & 0.1112 & 0.1014 & 0.0720 & 0.1005 & \textbf{0.0681} \\
\textit{CLaMP 3$_{saas}$}  & \textbf{0.1788} & \textbf{0.1980} & \textbf{0.1962} & \textbf{0.1665} & \textbf{0.1459} & \textbf{0.1770} & \textbf{0.1736} & \textbf{0.0945} & \textbf{0.1561} & 0.0675 \\
\bottomrule
\end{tabular}
\label{tab:multilingual_comparison}
\end{table*}

Beyond symbolic music retrieval, CLaMP 3 also achieved notable performances in audio retrieval. Both variants—CLaMP 3\(_{\textnormal {sa}}^{\textnormal {c2}}\) and CLaMP 3\(_{\textnormal {saas}}\)—consistently outperformed CLAP, with CLaMP 3\(_{\textnormal {saas}}\) standing out. It achieved the highest MRR of 0.1985 on SDD, marking a substantial improvement over TTMR++ (0.1437) and CLAP (0.1310). While TTMR++ performed well on MC-R (0.1248), its results on the original MusicCaps dataset are abnormally higher (see Table~\ref{tab:musiccaps_leakage}), likely because it was trained on half of MusicCaps’ original music-text pairs. This training overlap suggests that indirect data leakage affects its performance, even when evaluated on MC-R.

CLaMP 3's strong performance extends to WikiMT-X, with both variants outperforming baselines across all four semantic categories. In \textit{Background} and \textit{Analysis}, where texts provide rich cultural or technical details, CLaMP 3\(_{\textnormal {sa}}^{\textnormal {c2}}\) and CLaMP 3\(_{\textnormal {saas}}\) excelled, achieving MRRs of 0.4028 and 0.3382 (sheet music) and 0.2017 and 0.1711 (audio). \textit{Description} and \textit{Scene}, however, are much harder to retrieve because they are less specific and semantically sparse. \textit{Description} excludes explicit identifiers like titles or artist names, while \textit{Scene} focuses on abstract, visualized scenario depictions (rather than the music itself), both of which make retrieval more difficult. Even so, CLaMP 3 performed notably better, with CLaMP 3\(_{\textnormal {saas}}\) scoring 0.0988 (\textit{Description}) and 0.0963 (\textit{Scene}) in audio, compared to TTMR++ (0.0584, 0.0301). This improvement stems from M4-RAG's diverse annotations, which better equip CLaMP 3 to retrieve abstract, semantically sparse texts compared to baseline models trained on less diverse data.

\subsection{Multilingual Text-to-Music Retrieval}
Currently, no non-English music-text benchmarks exist, making multilingual evaluation challenging. To address this, we used SeamlessM4T \cite{barrault2023seamlessm4t} to translate WikiMT-X background annotations into multiple languages. To account for translation noise, BLEU scores \cite{DBLP:conf/acl/PapineniRWZ02} were calculated by comparing original texts with back-translations. The translated annotations were then used for retrieval of matching ABC notation, MIDI (from ABC), and audio files.

We carefully selected ten languages to ensure diversity in linguistic families, scripts, regions, and resource levels. Five UN official languages were chosen from those included in M4-RAG as they represent different cultures and regions with global significance. The other five, marked with asterisks in Table~\ref{tab:multilingual_comparison}, come from different linguistic families with distinct scripts and minimal vocabulary overlap, specifically to test CLaMP 3’s generalization to languages unseen in music-text alignment.

To the best of our knowledge, apart from CLaMP 3, CLaMP 2 is the only multilingual MIR model, but it is limited to symbolic music. No baselines exist for multilingual audio retrieval, as models like CLAP and TTMR++ are restricted to English.

CLaMP 3’s two variants differ in their language exposure. CLaMP 3\(_{\textnormal {sa}}^{\textnormal {c2}}\) initializes its text and symbolic music encoders from CLaMP 2, which was pre-trained on symbolic-text alignment across all XLM-R-supported languages, giving it prior exposure to all languages in Table~\ref{tab:multilingual_comparison}. In contrast, CLaMP 3\(_{\textnormal {saas}}\) has never aligned music data with the languages marked with asterisks, demonstrating true cross-lingual generalization in its performance.

Table~\ref{tab:multilingual_comparison} shows that CLaMP 3 demonstrates strong cross-lingual generalization in both symbolic music and audio retrieval tasks. For symbolic music retrieval, CLaMP 3\(_{\textnormal {sa}}^{\textnormal {c2}}\) clearly outperforms CLaMP 2 on all languages, including those not in M4-RAG, showing that full language coverage during training is not necessary for improved multilingual retrieval. Meanwhile, CLaMP 3\(_{\textnormal {saas}}\), without any prior alignment between these languages and music or specific optimization for symbolic music tasks, matches CLaMP 2’s performance on MIDI and surpasses it on ABC notation. This indicates that CLaMP 3\(_{\textnormal {saas}}\) achieves true cross-lingual generalization on unseen languages.

In audio retrieval, CLaMP 3\(_{\textnormal {saas}}\) performed well on languages it had never seen during alignment. For instance, it outperformed CLaMP 3\(_{\textnormal {sa}}^{\textnormal {c2}}\) on Finnish (0.1770 vs. 0.1112), Greek (0.1736 vs. 0.1014), and Kazakh (0.1561 vs. 0.1005), even though CLaMP 3\(_{\textnormal {sa}}^{\textnormal {c2}}\) had indirect exposure to these languages during CLaMP 2 pre-training. Notably, even for its weakest unseen language, Amharic (0.0675), CLaMP 3\(_{\textnormal {saas}}\) outperformed CLAP’s performance on English text (0.0598). This suggests that prior exposure to a language is not necessary for achieving strong audio retrieval performance.

The ability to retrieve languages beyond the training data stems from XLM-R’s cross-lingual semantics and the universal representations of CLaMP 3’s music encoders. This enables the model to handle low-resource languages and even generalize to unseen ones, enhancing its inclusivity and versatility for global MIR.

\begin{table}[t]
\centering
\renewcommand{\arraystretch}{1.25}
\setlength{\tabcolsep}{2pt} 
\fontsize{8.6}{9}\selectfont
\caption{Results for emergent cross-modal retrieval on WikiMT-X pairings across different musical modalities. \textbf{S}: Sheet Music (ABC notation), \textbf{P}: Performance Signals (MIDI, converted from ABC), \textbf{A}: Audio recordings.}
\begin{tabular}{l c c c c c c}
\toprule
\textbf{Model} & \textbf{S$\rightarrow$P} & \textbf{S$\rightarrow$A} & \textbf{P$\rightarrow$S} & \textbf{P$\rightarrow$A} & \textbf{A$\rightarrow$S} & \textbf{A$\rightarrow$P} \\
\midrule
\textit{CLaMP 2}        & \textbf{0.5138} & -      & 0.4480 & -      & -      & -      \\
\textit{CLaMP 3$_{sa}^{c2}$} & 0.4547 & 0.0543 & \textbf{0.5293} & 0.0313 & \textbf{0.0492} & \textbf{0.0383} \\
\textit{CLaMP 3$_{saas}$} & 0.3262 & \textbf{0.0578} & 0.3146 & \textbf{0.0397} & 0.0410 & 0.0303 \\
\bottomrule
\end{tabular}
\label{tab:emergent_cross_modal}
\end{table}

\subsection{Emergent Cross-Modal Retrieval}
Emergent cross-modal retrieval assesses a model’s ability to align and retrieve musical content across modalities without explicit alignment training, showcasing its capacity to generalize to unaligned modalities. Table~\ref{tab:emergent_cross_modal} reports results for all possible retrieval directions between ABC notation, MIDI, and audio data.

CLaMP 3 significantly advances cross-modal retrieval by supporting both symbolic and audio modalities, addressing a key limitation of CLaMP 2. While CLaMP 2 excels in symbolic tasks (S$\rightarrow$P: 0.5138, P$\rightarrow$S: 0.4480) without explicit alignment between ABC and MIDI, it cannot retrieve between symbolic and audio modalities.

In contrast, CLaMP 3\(_{\textnormal {sa}}^{\textnormal {c2}}\) not only achieves state-of-the-art performance on symbolic music tasks like P$\rightarrow$S (0.5293) but also enables emergent retrieval between symbolic music and audio. Similarly, CLaMP 3\(_{\textnormal {saas}}\), optimized for audio retrieval, achieves meaningful results on new tasks such as S$\rightarrow$A (0.0578) and P$\rightarrow$A (0.0397), demonstrating its ability to unify symbolic and audio modalities in a shared representation space.

While audio retrieval is inherently more challenging due to the continuous nature of audio signals, all directions achieve MRR scores well above the random baseline of 0.0075. Nonetheless, further optimization is required to reduce the performance gap between symbolic and audio retrieval.

\section{Conclusions}
In this paper, we introduced CLaMP 3, a unified MIR framework that aligns sheet music, performance signals, audio, and multilingual text using contrastive learning. CLaMP 3 demonstrates strong cross-modal and cross-lingual generalization, effectively handling unaligned modalities and unseen languages during training.

To address the lack of high-quality datasets, we curated M4-RAG, a collection of 2.31 million music-text pairs spanning 27 languages and 194 countries. We also released WikiMT-X, the first benchmark combining text, sheet music, and audio for comprehensive evaluation.

Our experiments show that CLaMP 3 achieves state-of-the-art performance in both symbolic and audio retrieval, excels in multilingual tasks, and enables retrieval across unaligned musical modalities. These results demonstrate its flexibility and the effectiveness of its shared representation space.

To conclude, CLaMP 3 sets a new standard in multimodal and multilingual MIR, demonstrating robust cross-modal and cross-lingual generalization. By releasing the CLaMP 3 model, M4-RAG dataset, and WikiMT-X benchmark, we provide resources to support future research in MIR and music generation across languages and modalities.

\section{Limitations}
Although CLaMP 3 attains state-of-the-art performance across modalities and languages, showing cross-modal and cross-lingual generalization, this work has several limitations that need to be addressed for further advancements in MIR.

First, while contrastive learning has advanced multimodal information retrieval, it struggles to capture the temporal dynamics of music. This is because such models typically use a single global representation to store the entire semantic content of a piece of music, making them insensitive to temporal dynamics. For example, in Beethoven's Symphony No. 5, the iconic four-note motif develops throughout the piece, yet current systems often miss this context. Addressing this requires moving beyond contrastive learning to incorporate temporal modeling, enabling systems to better capture nuances and deliver more context-aware and accurate retrieval.

Second, although Table~\ref{tab:multilingual_comparison} indicates that while CLaMP 3 can generalize to languages beyond music-text alignment, the multilingual text-to-music retrieval evaluation in it heavily relies on translation models due to the lack of native multilingual benchmarks. The translation quality varies significantly across languages, which introduces noise and reduces the reliability of evaluations. Developing native multilingual benchmarks is the primary and almost indispensable solution to achieve more accurate and fair assessments of model performance.

Finally, as shown in Table~\ref{tab:emergent_cross_modal}, the alignment between audio and symbolic modalities, though showing emergent capabilities with performance far above random, remains relatively weak. Addressing this limitation requires collecting paired data for supervised alignment and leveraging text as a bridging modality to further enhance connections between different musical modalities.

\section*{Authors}
Shangda Wu\textsuperscript{1}, \textit{shangda@mail.ccom.edu.cn}

\noindent
Zhancheng Guo\textsuperscript{1}, \textit{23a053@mail.ccom.edu.cn}

\noindent
Ruibin Yuan\textsuperscript{2}, \textit{ryuanab@connect.ust.hk}

\noindent
Junyan Jiang\textsuperscript{3, 4}, \textit{jj2731@nyu.edu} 

\noindent
Seungheon Doh\textsuperscript{5}, \textit{seungheondoh@kaist.ac.kr}

\noindent
Gus Xia\textsuperscript{3, 4}, \textit{Gus.Xia@mbzuai.ac.ae}

\noindent
Juhan Nam\textsuperscript{5}, \textit{juhan.nam@kaist.ac.kr}

\noindent
Xiaobing Li\textsuperscript{1}, \textit{lxiaobing@ccom.edu.cn}

\noindent
Feng Yu\textsuperscript{1}, \textit{yufengai@ccom.edu.cn}

\section*{Correspondence}
Maosong Sun\textsuperscript{1, 6}, \textit{sms@tsinghua.edu.cn}

\section*{Affiliations}
\textsuperscript{1}Central Conservatory of Music \\ 
\textsuperscript{2}Hong Kong University of Science and Technology \\ 
\textsuperscript{3}New York University Shanghai \\ 
\textsuperscript{4}Mohamed bin Zayed University of Artificial Intelligence \\
\textsuperscript{5}Korea Advanced Institute of Science and Technology \\
\textsuperscript{6}Tsinghua University

\section*{Acknowlegdements}
This work was supported by the following funding sources: Special Program of National Natural Science Foundation of China (Grant No. T2341003), Advanced Discipline Construction Project of Beijing Universities, Major Program of National Social Science Fund of China (Grant No. 21ZD19), and the National Culture and Tourism Technological Innovation Engineering Project (Research and Application of 3D Music).

In addition, we thank Flaticon\footnote{\url{https://www.flaticon.com}} for icons used in Fig.~\ref{fig:overview} and Fig.~\ref{fig:clamp3}, Yusong Wu (University of Montreal) for helping us understand CLAP in detail, and Monan Zhou (Central Conservatory of Music) for assisting with WikiMT-X data processing.

\bibliography{custom}

\appendix
\onecolumn
\section{Prompt and Examples}
\label{sec:prompt}

\vspace{1em} 
\noindent\makebox[\textwidth]{%
    \includegraphics[width=\textwidth]{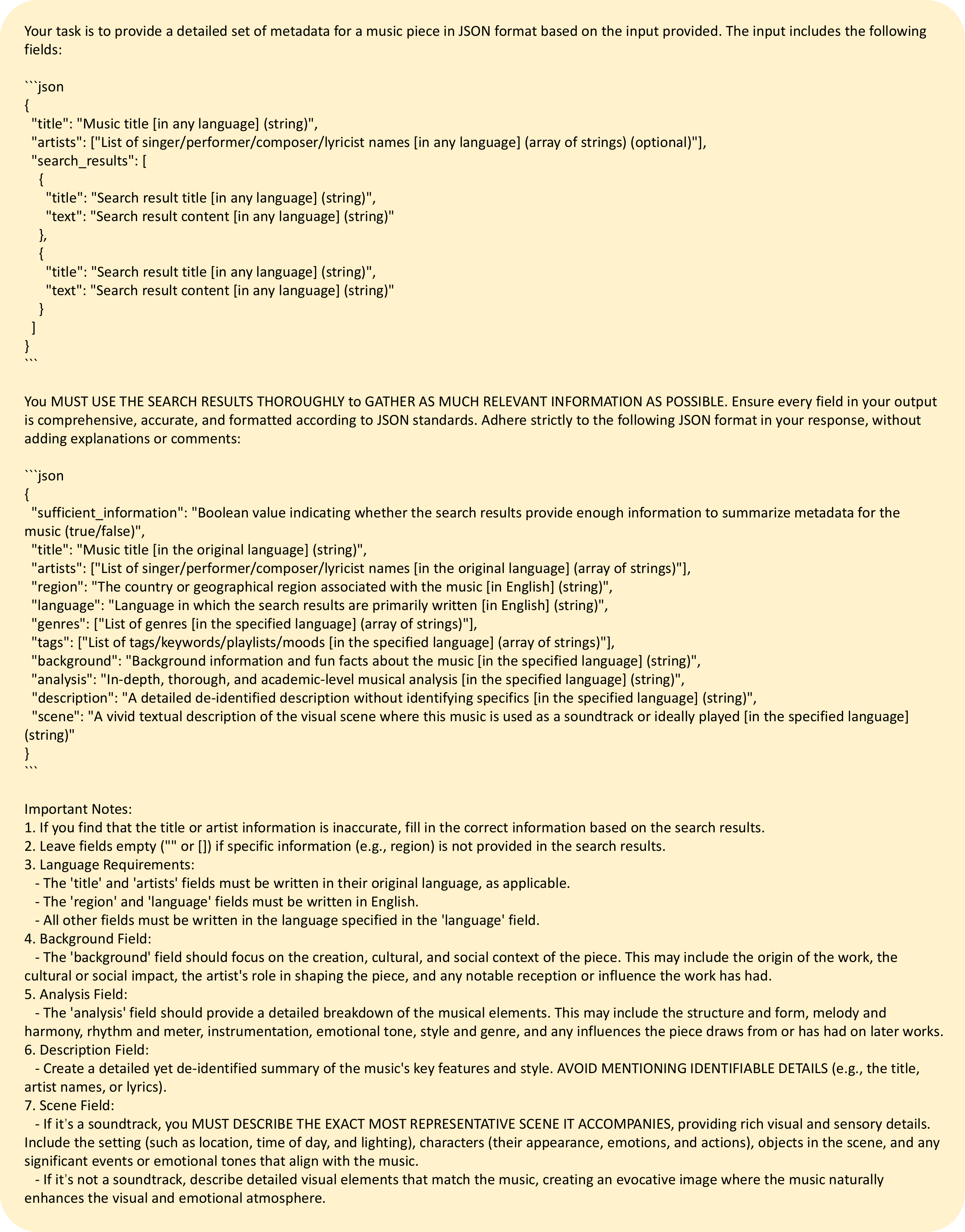}}
\captionof{figure}{The metadata generation prompt was used for constructing the M4-RAG dataset. This prompt outlines the required JSON structure for describing music metadata comprehensively, including fields for title, artists, region, language, genres, tags, background context, musical analysis, general description, and visual scene. Detailed instructions and formatting requirements are provided to ensure high-quality and consistent metadata extraction from search results. Based on our experience, we recommend adding the requirement to the prompt that \textit{Region} and \textit{Language} be output in accordance with ISO standards, which can reduce the need for post-processing.}

\clearpage

\vspace{1em} 
\noindent\makebox[\textwidth]{%
    \includegraphics[width=\textwidth]{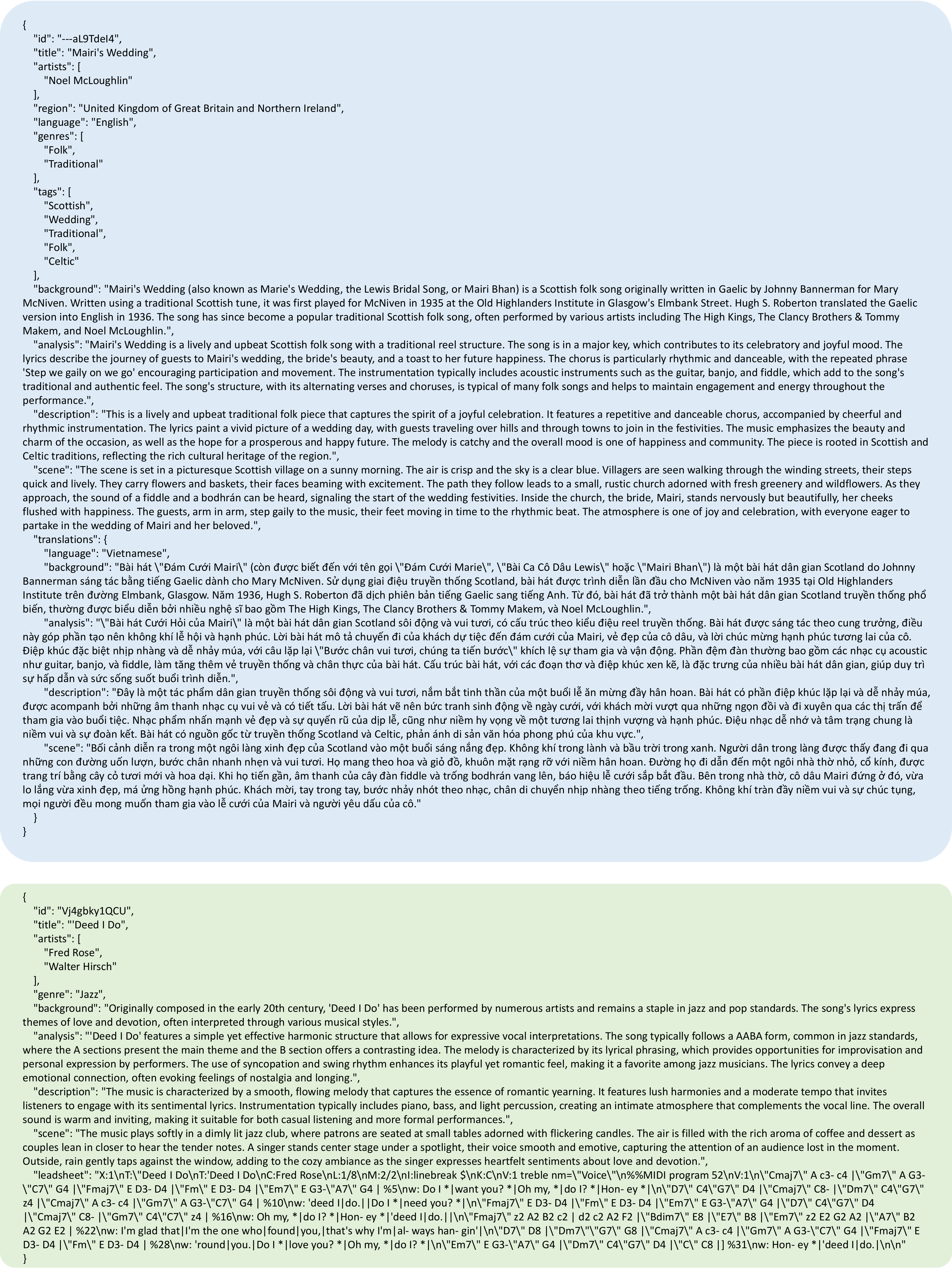}}
\captionof{figure}{Metadata examples from the M4-RAG and WikiMT-X datasets. The top section shows an entry for “Mairi's Wedding” from the M4-RAG dataset, including detailed multilingual metadata in English and Vietnamese, and an associated audio recording identified by a YouTube ID. The bottom section presents an entry for “Deed I Do” from the WikiMT-X dataset, which includes a YouTube ID linking to an audio recording, a genre label (Jazz, one of eight predefined categories), four types of long-form text annotations, and a lead sheet in ABC notation.}

\twocolumn
\setlength{\parindent}{2em}
\twocolumn[{
\noindent
\begin{minipage}{\textwidth}
    \centering
    \renewcommand{\arraystretch}{1.25}
    \setlength{\tabcolsep}{5pt}
    \fontsize{10}{9}\selectfont
    \captionof{table}{Average cosine similarity between text and music features across datasets and annotation types, alongside human ratings of music-text alignment and musical aesthetics. Cosine similarity reflects pairing quality as estimated by text-to-music retrieval models, while human ratings assess how well the text semantically aligns with the music (\textbf{Alignment}) and the perceived aesthetic quality of the music (\textbf{Aesthetics}).}
    \label{tab:eval_benchmark}
    \begin{tabular}{llcccccc}
    \toprule
    \textbf{Dataset} & \textbf{Annotation} & \textbf{CLaMP} & \textbf{CLaMP 2} & \textbf{CLaMP 3$_{\textnormal {sa}}^{\textnormal {c2}}$} & \textbf{CLaMP 3$_{\textnormal {saas}}$} & \textbf{Alignment} & \textbf{Aesthetics} \\
    \midrule
    \textit{WikiMT}   & \textit{Caption}           & 0.1900 & 0.1244 & 0.2028 & 0.2184 & \textbf{4.83} & 3.42 \\
    \textit{MidiCaps} & \textit{Caption}           & 0.1133 & 0.0255 & 0.1401 & 0.1583 & 3.92 & 2.83 \\
    \multirow{4}{*}{\textit{WikiMT-X}}
             & \textit{Background}  & \textbf{0.2429} & 0.1343 & 0.2239 & 0.2264 & 4.67 & \multirow{4}{*}{\textbf{3.50}} \\
             & \textit{Analysis}    & 0.1336 & 0.1097 & 0.2261 & 0.2461 & 3.75 & \\
             & \textit{Description} & 0.0794 & 0.0451 & \textbf{0.2359} & 0.2752 & 3.42 & \\
             & \textit{Scene}       & 0.0779 & \textbf{0.1410} & 0.2240 & \textbf{0.2874} & 3.67 & \\
    \midrule
    \midrule
    \textbf{Dataset} & \textbf{Annotation} & \textbf{CLAP} & \textbf{TTMR++} & \textbf{CLaMP 3$_{\textnormal {sa}}^{\textnormal {c2}}$} & \textbf{CLaMP 3$_{\textnormal {saas}}$} & \textbf{Alignment} & \textbf{Aesthetics}  \\
    \midrule
    \textit{SDD}    & \textit{Caption}           & 0.3446 & 0.3340 & \textbf{0.1129} & 0.1683 & 2.25 & 2.50 \\
    \textit{MusicCaps}   & \textit{Caption}           & 0.2518 & 0.3957 & 0.1077 & 0.1500 & 4.08 & 2.58 \\
    \multirow{4}{*}{\textit{WikiMT-X}}
           & \textit{Background}  & \textbf{0.3734} & \textbf{0.4477} & 0.0092 & 0.1186 & \textbf{4.50} & \multirow{4}{*}{\textbf{3.58}} \\
           & \textit{Analysis}    & 0.2594 & 0.3813 & 0.0024 & 0.1298 & 3.92 & \\
           & \textit{Description} & 0.2000 & 0.3340 & 0.0385 & 0.1738 & 3.33 & \\
           & \textit{Scene}       & 0.1848 & 0.2590 & 0.0525 & \textbf{0.1996} & 3.83 & \\
    \bottomrule
    \end{tabular}
    \vspace{2em}
\end{minipage}
}]

\section{Evaluation of Annotation Quality Across Benchmarks}
\label{sec:eval_benchmark}

To evaluate the semantic quality of music-text pairings, we conduct an evaluation combining (i) automatic similarity scores from models and (ii) human ratings on music-text alignment and musical aesthetics. The results are summarized in Table~\ref{tab:eval_benchmark}.

We include both LLM-generated and human-annotated datasets in this evaluation. Specifically, all symbolic datasets—WikiMT, MidiCaps, and WikiMT-X—contain text annotations generated by large language models. In contrast, the audio datasets differ in annotation quality: MusicCaps captions are written by trained musicians, while SDD was annotated by non-expert crowd workers.

For the automatic evaluation, we report the average cosine similarity between text and music embeddings generated by multiple retrieval models, including CLAP, TTMR++, CLaMP, CLaMP 2, and two CLaMP 3 variants. These similarity scores indicate how well the text and music are aligned in the shared embedding space.

WikiMT-X consistently achieves higher similarity scores across most annotation types compared to previous datasets. For example, its \textit{Background} annotations reach cosine similarities of 0.2429 (CLaMP), 0.3734 (CLAP), and 0.4477 (TTMR++), outperforming scores seen in other datasets. Similar improvements are observed in the \textit{Analysis} and \textit{Description} categories, with the highest overall score of 0.2874 achieved by CLaMP 3$_\textnormal{saas}$ on the \textit{Scene} annotation type.

To complement the automatic metrics, we conducted a human evaluation. Four conservatory-trained musicians (each with over 10 years of formal experience) rated 144 music-text examples based on two criteria: semantic alignment (how well the text matches the music) and musical aesthetics (the perceived quality of the music), using a 1-5 scale.

Results show that WikiMT-X performs on par with or better than expert-annotated datasets. In the symbolic domain, its \textit{Background} annotations receive a high alignment score of 4.67—close to WikiMT's 4.83 and higher than MidiCaps' 3.92.

In the audio domain, WikiMT-X again demonstrates strong performance. Its alignment scores reach up to 4.50, and aesthetics up to 3.58—substantially outperforming SDD (2.25 / 2.50) and MusicCaps (4.08 / 2.58), which include more amateur or crowd-sourced material. The consistently higher aesthetic ratings for WikiMT and WikiMT-X likely stem from their inclusion of well-known Western popular music from the 20th century, in contrast to the less polished recordings found in other benchmarks.

Overall, the results support WikiMT-X as a high-quality benchmark for multimodal text-to-music retrieval. Its annotations show robust semantic alignment and musical relevance, confirmed by both model metrics and expert ratings. With careful filtering and validation, LLM-generated annotations can match or even exceed expert quality across both symbolic and audio datasets.

\clearpage

\section{\textit{t}-SNE Visualizations on WikiMT-X}
\label{sec:t-sne}
We apply \textit{t}-SNE (t-distributed Stochastic Neighbor Embedding) to the WikiMT-X dataset to visualize how CLaMP 3 organizes data into a shared representation space. The projections illustrate the model’s ability to align data across modalities, languages, and semantic categories.

Fig.~\ref{figure:t-sne,a} includes features from Text (background annotations), ABC notation, MIDI, and Audio. Each modality forms a distinct cluster, reflecting the inherent differences in how information is encoded. Notably, modalities closer to Text tend to perform better, aligning with the trend in Table~\ref{tab:multilingual_comparison}, suggesting a correlation between embedding proximity and cross-modal effectiveness. Additionally, all musical modalities display a mirrored symmetry around the Text cluster, indicating that Text may serve as a semantic anchor. This symmetry suggests CLaMP 3 aligns modalities relative to Text, balancing modality-specific features while preserving semantic consistency.

Fig.~\ref{figure:t-sne,b} focuses on background annotations in four languages—English, Spanish, Chinese, and Amharic—selected to represent varying retrieval performance levels. Despite their linguistic differences, these languages largely overlap, indicating strong cross-lingual alignment. English and Spanish cluster closely, reflecting both their shared linguistic roots. Chinese shows moderate overlap with English, suggesting that CLaMP 3 effectively bridges typologically distant languages. However, Amharic, a low-resource and unseen language, forms more isolated clusters, indicating the challenges of aligning low-resource languages.

Fig.~\ref{figure:t-sne,c} shows four semantic categories—\textit{Background}, \textit{Analysis}, \textit{Description}, and \textit{Scene}—presenting how CLaMP 3 handles different content types. \textit{Background}, \textit{Analysis}, and \textit{Description} often converge, reflecting the overlap in explanatory texts as they cover related musical concepts. In contrast, \textit{Scene} forms distinct clusters, likely because it focuses on visual depictions, leading to more consistent semantic patterns tied to specific imagery rather than music.

Across all three visualizations, genre boundaries remain clear despite differences in modality, language, or semantic category. This shows that CLaMP 3 effectively aligns multimodal and multilingual data while preserving genre-specific distinctions, demonstrating the model’s strong representational capabilities.

\begin{figure}[t]
    \centering
    \begin{subfigure}{0.48\textwidth}
        \centering
        \includegraphics[width=\textwidth]{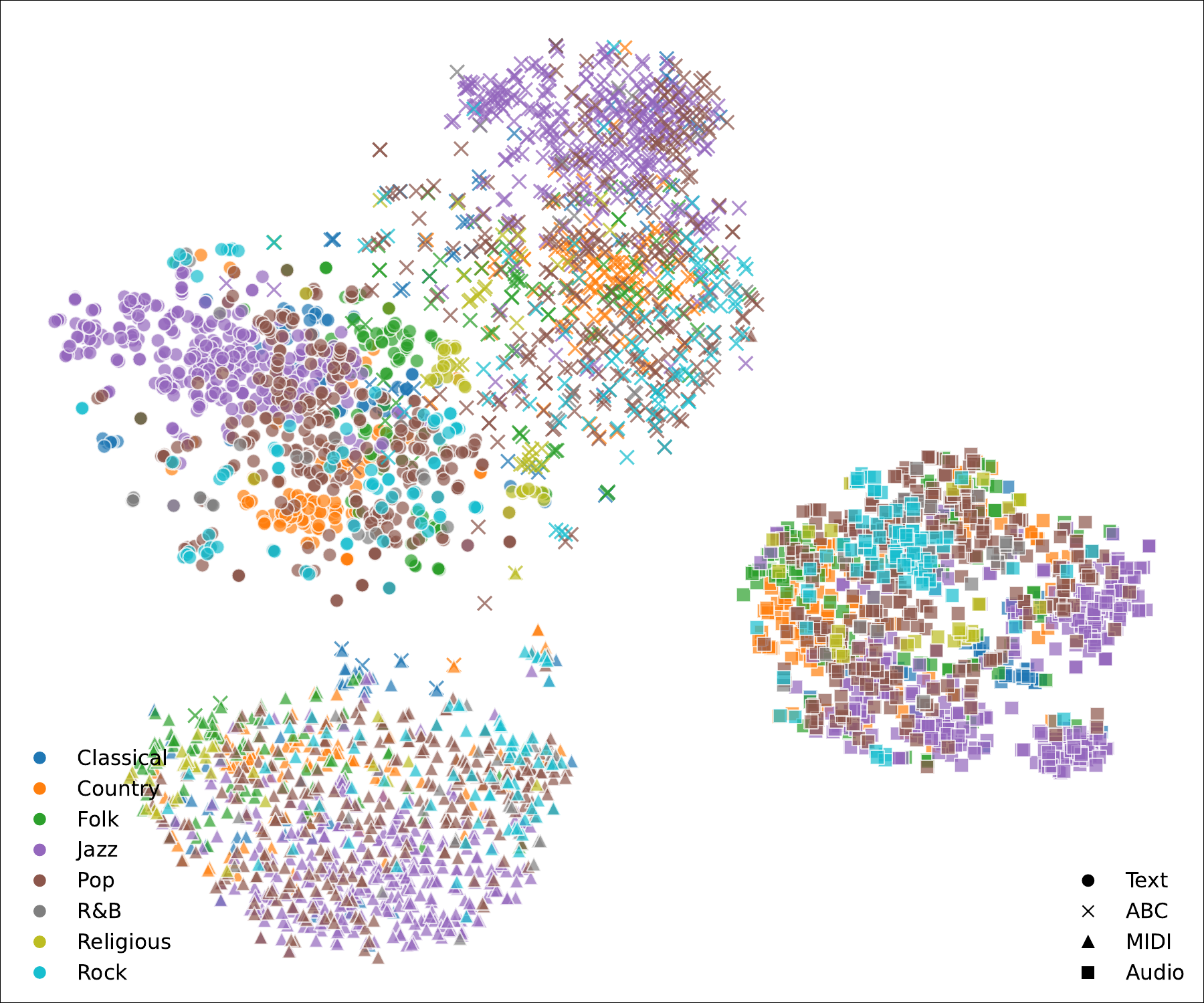}
        \caption{Modality}
        \vspace{1em}
        \label{figure:t-sne,a}
    \end{subfigure}
    \hfill
    \begin{subfigure}{0.48\textwidth}
        \centering
        \includegraphics[width=\textwidth]{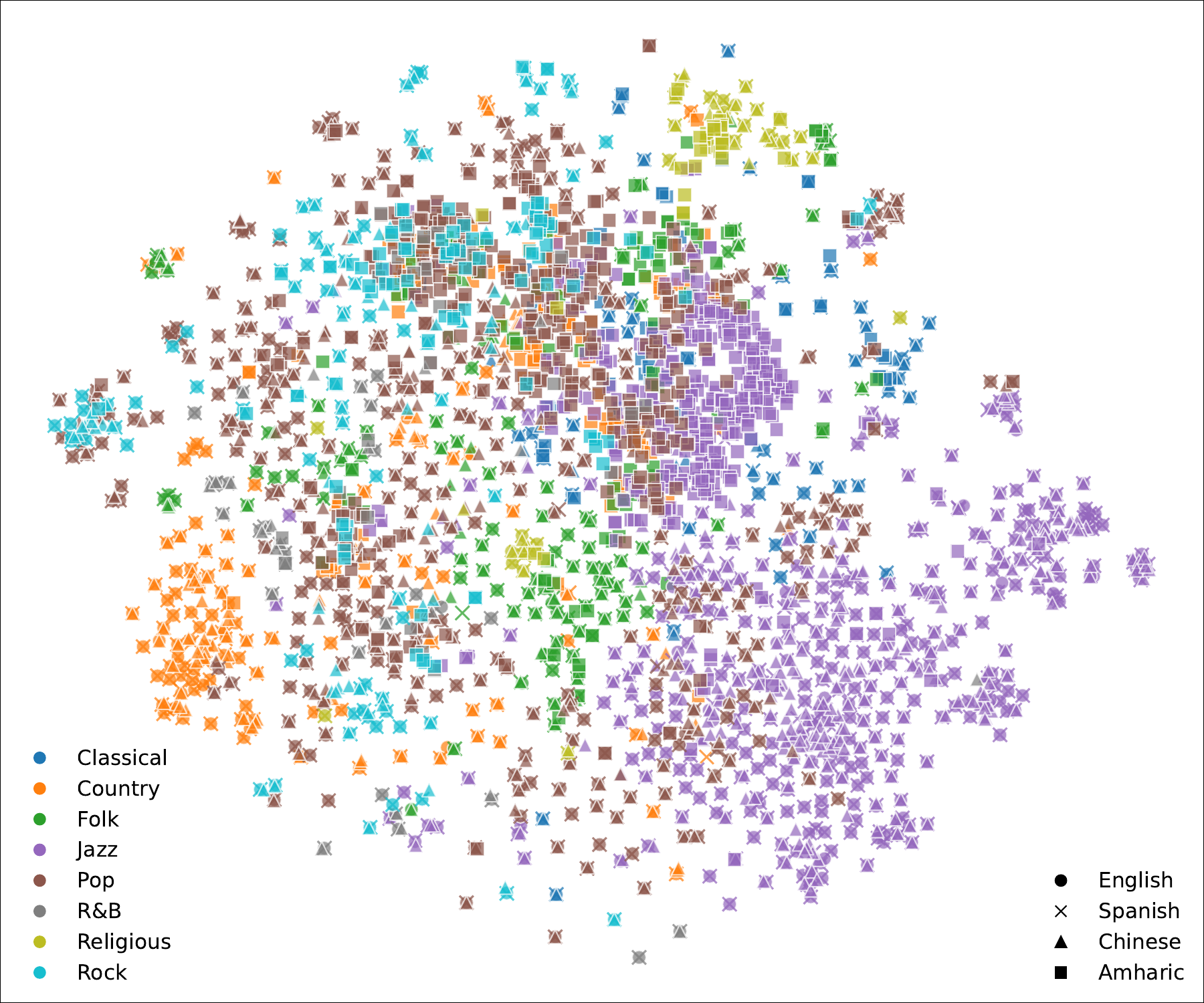}
        \caption{Language}
        \vspace{1em}
        \label{figure:t-sne,b}
    \end{subfigure}
    \begin{subfigure}{0.48\textwidth}
        \centering
        \includegraphics[width=\textwidth]{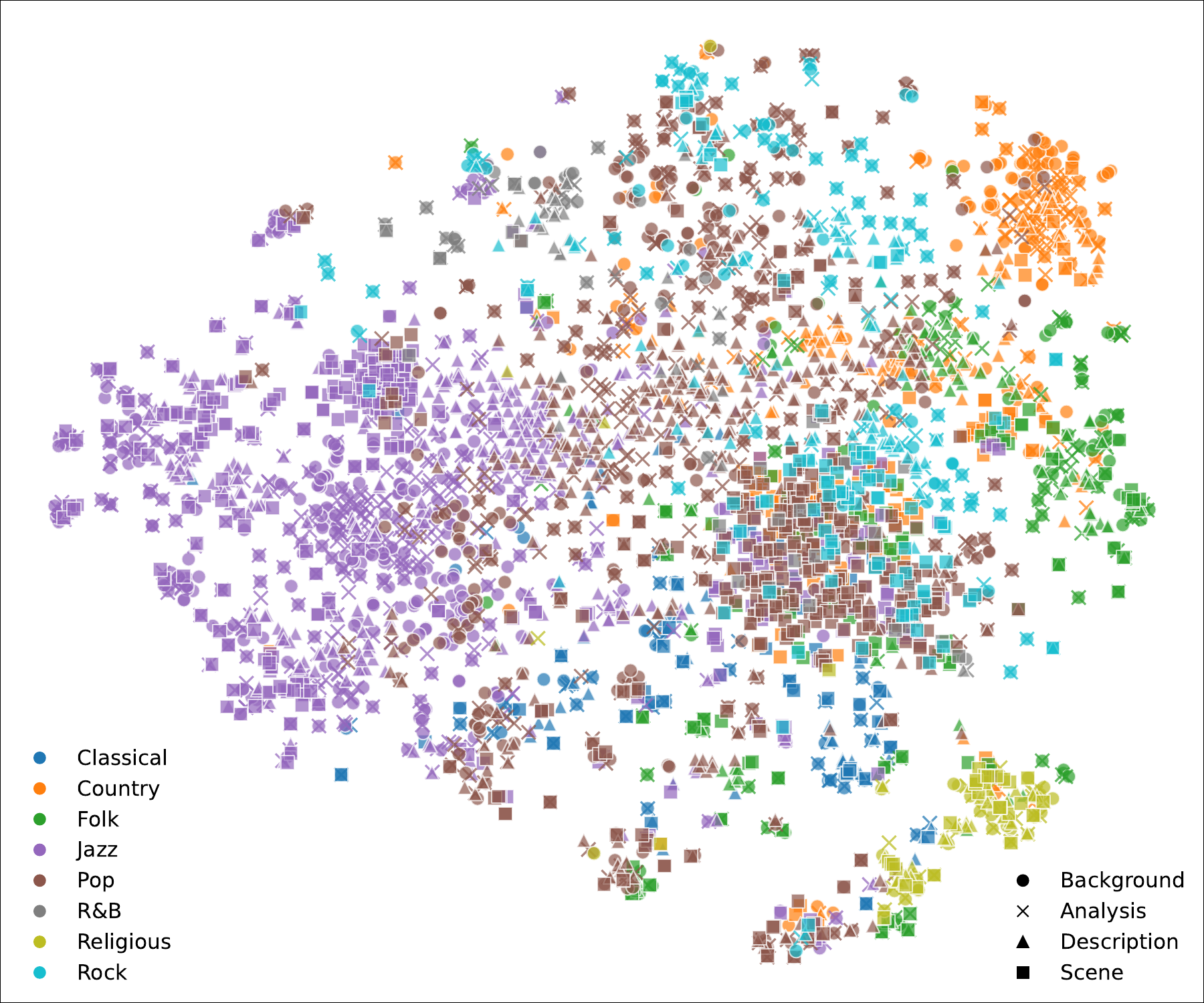}
        \caption{Semantics}
        \label{figure:t-sne,c}
    \end{subfigure}
    \caption{\textit{t}-SNE visualization of the WikiMT-X dataset, illustrating the distribution of samples based on three distinct factors: (a) Modality, (b) Language, and (c) Semantics. The representations are extracted using CLaMP 3\(_{\textnormal {saas}}\). Each point represents a data sample, colored according to its genre.}
\end{figure}

\clearpage

\begin{table*}[h]
\centering
\renewcommand{\arraystretch}{1.25}
\setlength{\tabcolsep}{12.5pt}
\fontsize{9.7}{9.5}\selectfont
\caption{Results for English text-to-music retrieval on several benchmarks: WikiMT and MidiCaps have 1,010 pairs, Song Describer Dataset (SDD) has 706 audio and 1,106 captions, and MusicCaps-Remake (MC-R) contains 2,777 pairs. MC-R prevents data leakage by using full-length audio and rewritten captions from AudioSet’s evaluation set.}
\begin{tabular}{l c c c c c c}
\toprule
\multirow{2}{*}{\textbf{\vspace{-2mm}Model}} 
& \multicolumn{2}{c}{\textbf{Symbolic Benchmarks}} 
& \multicolumn{4}{c}{\textbf{WikiMT-X (Sheet Music)}} \\ 
\cmidrule(lr){2-3} \cmidrule(lr){4-7}
& \textit{WikiMT} & \textit{MidiCaps} 
& \textit{Background} & \textit{Analysis} & \textit{Description} & \textit{Scene} \\
\midrule
\textit{CLaMP 3$_{as}$}   & 0.1973 & 0.0788 & 0.2108 & 0.1660 & 0.1049 & 0.1056 \\
\textit{CLaMP 3$_{sa}$}   & 0.3789 & 0.1322 & 0.3591 & 0.3088 & 0.1316 & \textbf{0.1643} \\
\textit{CLaMP 3$_{sa}^{c2}$}   & \textbf{0.4498} & \textbf{0.2826} & \textbf{0.4028} & \textbf{0.3382} & 0.0835 & 0.1512 \\
\textit{CLaMP 3$_{assa}$}  & 0.2993 & 0.0884 & 0.2919 & 0.2507 & \textbf{0.1459} & 0.1464 \\
\textit{CLaMP 3$_{saas}$}  & 0.3555 & 0.1798 & 0.3301 & 0.2758 & 0.1274 & 0.1512 \\
\textit{CLaMP 3$_{saas}^{c2}$}   & 0.3631 & 0.2688 & 0.3295 & 0.2957 & 0.0951 & 0.1395 \\
\midrule
\midrule
\multirow{2}{*}{\textbf{\vspace{-2mm}Model}} 
& \multicolumn{2}{c}{\textbf{Audio Benchmarks}} 
& \multicolumn{4}{c}{\textbf{WikiMT-X (Audio)}} \\ 
\cmidrule(lr){2-3} \cmidrule(lr){4-7}
& \textit{SDD} & \textit{MC-R} 
& \textit{Background} & \textit{Analysis} & \textit{Description} & \textit{Scene} \\
\midrule
\textit{CLaMP 3$_{as}$}   & 0.1977 & 0.1117 & 0.1602 & 0.1375 & 0.0854 & 0.0819 \\
\textit{CLaMP 3$_{sa}$}   & 0.1607 & 0.0937 & 0.1718 & 0.1586 & 0.0997 & 0.0871 \\
\textit{CLaMP 3$_{sa}^{c2}$}   & 0.1612 & 0.0959 & 0.1180 & 0.1206 & 0.0639 & 0.0619 \\
\textit{CLaMP 3$_{assa}$}  & 0.2003 & 0.1045 & 0.1597 & 0.1522 & \textbf{0.1020} & 0.0873 \\
\textit{CLaMP 3$_{saas}$}  & 0.1985 & 0.1177 & \textbf{0.2017} & \textbf{0.1711} & 0.0988 & \textbf{0.0963} \\
\textit{CLaMP 3$_{saas}^{c2}$}   & \textbf{0.2115} & \textbf{0.1180} & 0.1583 & 0.1530 & 0.0768 & 0.0885 \\
\bottomrule
\end{tabular}
\label{tab:en_variants}
\end{table*}

\begin{table*}[h]
\centering
\renewcommand{\arraystretch}{1.25}
\setlength{\tabcolsep}{6pt}
\fontsize{9.7}{9.5}\selectfont
\caption{Results for multilingual text-to-music retrieval on translated WikiMT-X background annotations. Languages marked with asterisks were not included in the M4-RAG training data. The BLEU scores below each language are calculated by back-translating the text with the SeamlessM4T model and comparing it to the original English text.}
\begin{tabular}{l c c c c c c c c c c}
\toprule
\multirow{2}{*}{\textbf{Model}} & \textbf{ru} & \textbf{fr} & \textbf{es} & \textbf{ar} & \textbf{zh} & \textbf{fi*} & \textbf{el*} & \textbf{ta*} & \textbf{kk*} & \textbf{am*} \\
& \textit{49.69} & \textit{55.50} & \textit{62.82} & \textit{53.38} & \textit{39.58} & \textit{39.19} & \textit{55.55} & \textit{40.07} & \textit{36.57} & \textit{56.08} \\ 
\midrule
\textbf{ABC Notation} & & & & & & & & & & \\
\textit{CLaMP 3$_{as}$}        & 0.1750 & 0.1931 & 0.1964 & 0.1594 & 0.1559 & 0.1828 & 0.1641 & 0.0997 & 0.1575 & 0.0876 \\
\textit{CLaMP 3$_{sa}$}        & 0.3262 & 0.3544 & 0.3536 & 0.3072 & \textbf{0.2459} & 0.3163 & 0.2879 & 0.1336 & 0.2894 & 0.1317 \\
\textit{CLaMP 3$_{sa}^{c2}$}   & \textbf{0.3614} & \textbf{0.3949} & \textbf{0.3921} & \textbf{0.3155} & 0.2373 & \textbf{0.3524} & \textbf{0.3226} & 0.1415 & \textbf{0.3397} & \textbf{0.1871} \\
\textit{CLaMP 3$_{assa}$}      & 0.2648 & 0.2810 & 0.2817 & 0.2450 & 0.2271 & 0.2644 & 0.2415 & \textbf{0.1432} & 0.2561 & 0.1300 \\
\textit{CLaMP 3$_{saas}$}      & 0.2918 & 0.3214 & 0.3239 & 0.2789 & 0.2358 & 0.2919 & 0.2681 & 0.1246 & 0.2703 & 0.1139 \\
\textit{CLaMP 3$_{saas}^{c2}$} & 0.2954 & 0.3171 & 0.3225 & 0.2773 & 0.2144 & 0.2990 & 0.2721 & 0.1348 & 0.2750 & 0.1690 \\
\midrule
\textbf{MIDI} & & & & & & & & & & \\
\textit{CLaMP 3$_{as}$}        & 0.0418 & 0.0416 & 0.0432 & 0.0404 & 0.0332 & 0.0456 & 0.0449 & 0.0297 & 0.0398 & 0.0267 \\
\textit{CLaMP 3$_{sa}$}        & 0.1174 & 0.1284 & 0.1316 & 0.1132 & 0.0890 & 0.1217 & 0.1112 & 0.0623 & 0.1117 & 0.0540 \\
\textit{CLaMP 3$_{sa}^{c2}$}   & \textbf{0.1921} & \textbf{0.2101} & \textbf{0.2137} & \textbf{0.1681} & \textbf{0.1316} & \textbf{0.2019} & \textbf{0.1702} & \textbf{0.0804} & \textbf{0.1765} & \textbf{0.1039} \\
\textit{CLaMP 3$_{assa}$}      & 0.0565 & 0.0582 & 0.0620 & 0.0582 & 0.0517 & 0.0620 & 0.0585 & 0.0394 & 0.0595 & 0.0354 \\
\textit{CLaMP 3$_{saas}$}      & 0.1165 & 0.1319 & 0.1330 & 0.1141 & 0.0937 & 0.1245 & 0.1143 & 0.0601 & 0.1104 & 0.0544 \\
\textit{CLaMP 3$_{saas}^{c2}$} & 0.1499 & 0.1645 & 0.1664 & 0.1408 & 0.1049 & 0.1560 & 0.1399 & 0.0653 & 0.1335 & 0.0793 \\
\midrule
\textbf{Audio} & & & & & & & & & & \\
\textit{CLaMP 3$_{as}$}        & 0.1267 & 0.1515 & 0.1525 & 0.1210 & 0.1089 & 0.1430 & 0.1428 & 0.0610 & 0.1043 & 0.0559 \\
\textit{CLaMP 3$_{sa}$}        & 0.1619 & 0.1717 & 0.1714 & 0.1529 & 0.1414 & 0.1585 & 0.1544 & \textbf{0.0991} & 0.1456 & 0.0774 \\
\textit{CLaMP 3$_{sa}^{c2}$}   & 0.1068 & 0.1150 & 0.1202 & 0.0981 & 0.0877 & 0.1112 & 0.1014 & 0.0720 & 0.1005 & 0.0681 \\
\textit{CLaMP 3$_{assa}$}      & 0.1426 & 0.1580 & 0.1588 & 0.1370 & 0.1202 & 0.1468 & 0.1431 & 0.0795 & 0.1276 & 0.0617 \\
\textit{CLaMP 3$_{saas}$}      & \textbf{0.1788} & \textbf{0.1980} & \textbf{0.1962} & \textbf{0.1665} & \textbf{0.1459} & \textbf{0.1770} & \textbf{0.1736} & 0.0945 & \textbf{0.1561} & 0.0675 \\
\textit{CLaMP 3$_{saas}^{c2}$} & 0.1331 & 0.1566 & 0.1554 & 0.1304 & 0.1208 & 0.1550 & 0.1460 & 0.0901 & 0.1340 & \textbf{0.0874} \\
\bottomrule
\end{tabular}
\label{tab:multilingual_variants}
\end{table*}

\clearpage

\begin{table}[t]
\centering
\renewcommand{\arraystretch}{1.25}
\setlength{\tabcolsep}{2pt}
\fontsize{8.6}{10}\selectfont
\caption{Results for emergent cross-modal retrieval on WikiMT-X pairings across different musical modalities. \textbf{S}: Sheet Music (ABC notation), \textbf{P}: Performance Signals (MIDI, converted from ABC), \textbf{A}: Audio recordings.}
\begin{tabular}{l c c c c c c}
\toprule
\textbf{Model} & \textbf{S$\rightarrow$P} & \textbf{S$\rightarrow$A} & \textbf{P$\rightarrow$S} & \textbf{P$\rightarrow$A} & \textbf{A$\rightarrow$S} & \textbf{A$\rightarrow$P} \\
\midrule
\textit{CLaMP 3$_{as}$}        & 0.1637 & 0.0557 & 0.1477 & 0.0248 & 0.0456 & 0.0237 \\
\textit{CLaMP 3$_{sa}$}        & 0.3205 & \textbf{0.0739} & 0.3054 & 0.0397 & 0.0479 & 0.0237 \\
\textit{CLaMP 3$_{sa}^{c2}$}   & \textbf{0.4547} & 0.0543 & \textbf{0.5293} & 0.0313 & 0.0492 & 0.0383 \\
\textit{CLaMP 3$_{assa}$}      & 0.1911 & 0.0619 & 0.1646 & 0.0299 & 0.0513 & 0.0264 \\
\textit{CLaMP 3$_{saas}$}      & 0.3262 & 0.0578 & 0.3146 & 0.0397 & 0.0410 & 0.0303 \\
\textit{CLaMP 3$_{saas}^{c2}$} & 0.3909 & 0.0688 & 0.4375 & \textbf{0.0467} & \textbf{0.0558} & \textbf{0.0431} \\
\bottomrule
\end{tabular}
\label{tab:emergent_variants}
\end{table}

\section{Performance of CLaMP 3 Variants}
\label{sec:variants}
A straightforward way to train CLaMP 3 would be to align symbolic music, audio, and text all at once. However, early experiments showed that this led to unstable training. The text encoder struggled because symbolic and audio data had very different distributions (Fig.\ref{figure:t-sne,a}) and pulled it in opposite directions, making alignment ineffective. To solve this, we adopted a multi-stage alignment strategy (Sec.~\ref{sec:training}) that gradually integrates each modality, ensuring stable and effective alignment.

To explore the best way to align modalities, we tested different training orders, leading to several model variants. The main difference among them is how and when the text encoder is aligned with symbolic music and audio encoders:

\textbf{CLaMP 3\(_{\textnormal {as}}\):} A two-stage alignment where text is first aligned with audio, then the text encoder is frozen while aligning with symbolic music.

\textbf{CLaMP 3\(_{\textnormal {sa}}\):} The reverse of CLaMP 3\(_{\textnormal {as}}\), first aligning text with symbolic music, then freezing the text encoder while aligning with audio.

\textbf{CLaMP 3\(_{\textnormal {sa}}^{\textnormal {c2}}\):} Same as CLaMP 3\(_{\textnormal {sa}}\), but starting with pre-trained text and symbolic encoders from CLaMP 2.

\textbf{CLaMP 3\(_{\textnormal {assa}}\):} A four-stage alignment: audio $\rightarrow$ symbolic $\rightarrow$ symbolic $\rightarrow$ audio, with the text encoder frozen in the second and fourth stages to maintain stability.

\textbf{CLaMP 3\(_{\textnormal {saas}}\):} A four-stage alignment: symbolic $\rightarrow$ audio $\rightarrow$ audio $\rightarrow$ symbolic, also freezing the text encoder in the second and fourth stages.

\textbf{CLaMP 3\(_{\textnormal {saas}}^{\textnormal {c2}}\):} Same as CLaMP 3\(_{\textnormal {saas}}\), but initialized with pre-trained text and symbolic encoders from CLaMP 2.

We evaluate these six variants across all experiments in Sec.~\ref{sec:exp} to assess their effectiveness in different retrieval tasks.

Table~\ref{tab:en_variants} shows that aligning text with symbolic music before audio improves generalization in English text-to-music retrieval. CLaMP 3\(_{\textnormal{sa}}\) outperforms CLaMP 3\(_{\textnormal{as}}\) in symbolic retrieval without compromising audio performance. Four-stage models outperform two-stage models in audio retrieval, emphasizing the importance of iterative alignment. Among them, CLaMP 3\(_{\textnormal{saas}}\) achieves the best balance between symbolic and audio retrieval. Leveraging CLaMP 2’s weight initialization enhances symbolic retrieval, as seen in CLaMP 3\(_{\textnormal{sa}}^{\textnormal{c2}}\) leading symbolic tasks. However, it does not consistently improve audio retrieval, likely because CLaMP 2 was trained only on symbolic music, limiting its text encoder’s adaptability to audio alignment.

Table~\ref{tab:multilingual_variants} demonstrates the impact of pre-training and training order on multilingual text-to-music retrieval. In symbolic retrieval, using CLaMP 2’s pre-trained text-symbolic encoders provides a clear advantage, with CLaMP 3\(_{\textnormal{sa}}^{\textnormal{c2}}\) achieving the highest scores across most languages. This suggests that pre-training helps build a strong shared representation space, especially for MIDI, where M4-RAG’s limited native data weakens overall performance. However, pre-training is not always decisive, as some non-pretrained models surpass pre-trained variants in certain languages for ABC retrieval. In contrast, audio retrieval is consistently strongest with CLaMP 3\(_{\textnormal{saas}}\), even in unseen languages, suggesting that training order plays a more crucial role in cross-lingual generalization.

Table~\ref{tab:emergent_variants} evaluates emergent cross-modal retrieval, where no direct supervised alignment exists among musical modalities. CLaMP 3\(_{\textnormal{sa}}^{\textnormal{c2}}\) achieves the best symbolic retrieval (S$\leftrightarrow$P), showing that CLaMP 2 pre-training strengthens symbolic-text alignment, which indirectly benefits symbolic retrieval. For symbolic-audio retrieval, CLaMP 3\(_{\textnormal{saas}}^{\textnormal{c2}}\) performs best, leading in P$\rightarrow$A (0.0467), A$\rightarrow$S (0.0558), and A$\rightarrow$P (0.0431). It consistently outperforms CLaMP 3\(_{\textnormal{saas}}\), suggesting that pre-training provides a stronger shared representation space, leading to better cross-modal generalization between unpaired modalities.

These results show the importance of both training order and pre-training in MIR. Multi-stage alignment stabilizes training, while training order plays a key role, particularly in audio retrieval and cross-lingual generalization. Pre-training with CLaMP 2 strengthens symbolic retrieval and improves cross-modal generalization, but its benefits are limited for audio retrieval.

\clearpage

\begin{table*}[t]
\centering
\renewcommand{\arraystretch}{1}
\setlength{\tabcolsep}{8.5pt}
\fontsize{9.7}{10}\selectfont
\caption{Symbolic classification performance for ABC notation and MIDI was assessed across three datasets: WikiMT (1,010 pieces, 8 genres), VGMIDI (204 pieces, 4 emotions), and Pianist8 (411 pieces, 8 composers).}
\begin{tabular}{l c c c c c c c}
\toprule
\multirow{2}{*}{\centering\textbf{\vspace{-2mm}Model}} & \multirow{2}{*}{\centering\textbf{\vspace{-2mm}Modality}} & \multicolumn{2}{c}{\textbf{WikiMT}} & \multicolumn{2}{c}{\textbf{VGMIDI}} & \multicolumn{2}{c}{\textbf{Pianist8}} \\
\cmidrule(lr){3-4} \cmidrule(lr){5-6} \cmidrule(lr){7-8}
& & \textit{F1-macro} & \textit{Accuracy} & \textit{F1-macro} & \textit{Accuracy} & \textit{F1-macro} & \textit{Accuracy} \\
\midrule
\textit{M3} & ABC & 0.2349 & 0.4010 & 0.6016 & 0.6341 & 0.7395 & 0.7590 \\
\textit{CLaMP} & ABC & 0.3452 & 0.4267 & 0.6453 & 0.6866 & 0.7067 & 0.7152 \\
\textit{CLaMP 2} & ABC & \textbf{0.3990} & \textbf{0.4653} & 0.7449 & \textbf{0.8049} & \textbf{0.8025} & \textbf{0.8072} \\
\textit{CLaMP 3$_{as}$} & ABC & 0.3135 & 0.4307 & 0.6638 & 0.7073 & 0.6872 & 0.6867 \\
\textit{CLaMP 3$_{sa}$} & ABC & 0.3225 & 0.4455 & 0.7725 & \textbf{0.8049} & 0.7403 & 0.7590 \\
\textit{CLaMP 3$_{sa}^{c2}$} & ABC & 0.3316 & 0.4356 & 0.6845 & 0.7317 & 0.7722 & 0.7711 \\
\textit{CLaMP 3$_{assa}$} & ABC & 0.3102 & 0.4455 & 0.4990 & 0.6341 & 0.6796 & 0.6988 \\
\textit{CLaMP 3$_{saas}$} & ABC & 0.3177 & 0.4356 & \textbf{0.7969} & \textbf{0.8049} & 0.7716 & 0.7952 \\
\textit{CLaMP 3$_{saas}^{c2}$} & ABC & 0.3568 & 0.4257 & 0.6694 & 0.7561 & 0.7891 & 0.7952 \\
\midrule
\midrule
\textit{M3} & MIDI & 0.2621 & 0.4257 & 0.5399 & 0.6098 & \textbf{0.9199} & \textbf{0.9157} \\
\textit{CLaMP 2} & MIDI & 0.2898 & 0.4455 & 0.5246 & 0.6585 & 0.8927 & 0.8916 \\
\textit{CLaMP 3$_{as}$} & MIDI & \textbf{0.3361} & \textbf{0.4653} & 0.5600 & 0.5854 & 0.8186 & 0.8313 \\
\textit{CLaMP 3$_{sa}$} & MIDI & 0.2614 & 0.4010 & \textbf{0.6864} & \textbf{0.7073} & 0.8461 & 0.8554 \\
\textit{CLaMP 3$_{sa}^{c2}$} & MIDI & 0.3073 & 0.4455 & 0.6223 & \textbf{0.7073} & 0.8696 & 0.8675 \\
\textit{CLaMP 3$_{assa}$} & MIDI & 0.2882 & 0.4406 & 0.5001 & 0.6098 & 0.8076 & 0.8193 \\
\textit{CLaMP 3$_{saas}$} & MIDI & 0.2721 & 0.4158 & 0.5723 & 0.6341 & 0.7834 & 0.7952 \\
\textit{CLaMP 3$_{saas}^{c2}$} & MIDI & 0.2943 & 0.4208 & 0.5474 & 0.6829 & 0.8565 & 0.8554 \\
\bottomrule
\end{tabular}
\label{tab:symbolic_cls}
\end{table*}

\begin{table*}[t]
\centering
\renewcommand{\arraystretch}{1}
\setlength{\tabcolsep}{4pt}
\fontsize{9.7}{11}\selectfont
\caption{Audio classification performance is evaluated on multiple benchmarks included in MARBLE: MTT (25,860 clips, 50 tags), GS (7,035 clips, 24 keys), GTZAN (1,000 clips, 10 genres), EMO (744 clips, valence/arousal regression), Nsynth (305,979 clips, 11 instrument categories, 88 pitches), and VocalSet (7,506 clips, 17 singing techniques, 20 singers).}
\begin{tabular}{l c c c c c c c c c c}
\toprule
\multirow{3}{*}{\textbf{\vspace{-3mm}Model}} & 
\multicolumn{2}{c}{\textbf{MTT}} & 
\multicolumn{1}{c}{\textbf{GS}} & 
\multicolumn{1}{c}{\textbf{GTZAN}} & 
\multicolumn{2}{c}{\textbf{EMO}} & 
\multicolumn{1}{c}{\textbf{Nsynth}} & 
\multicolumn{1}{c}{\textbf{Nsynth}} & 
\multicolumn{1}{c}{\textbf{VocalSet}} & 
\multicolumn{1}{c}{\textbf{VocalSet}} \\
& \multicolumn{2}{c}{Tagging} & Key & Genre & \multicolumn{2}{c}{Emotion} & Instrument & Pitch & Tech & Singer \\
\cmidrule(lr){2-3} \cmidrule(lr){4-4} \cmidrule(lr){5-5} \cmidrule(lr){6-7} 
\cmidrule(lr){8-8} \cmidrule(lr){9-9} \cmidrule(lr){10-10} \cmidrule(lr){11-11} 
& \textit{ROC} & \textit{AP} & \textit{Acc} & \textit{Acc} & \textit{R2$^V$} & \textit{R2$^A$} & \textit{Acc} & \textit{Acc} & \textit{Acc} & \textit{Acc} \\
\midrule
\textit{MERT$_{mean}$} & 0.9068 & 0.3915 & \textbf{0.6475} & 0.6689 & 0.5185 & \textbf{0.7501} & 0.6963 & \textbf{0.9152} & \textbf{0.7219} & \textbf{0.8961} \\
\textit{CLAP} & 0.9066 & 0.3897 & 0.1596 & 0.8207 & 0.5408 & 0.7025 & \textbf{0.7817} & 0.5146 & 0.6868 & 0.6327 \\
\textit{TTMR++} & 0.9082 & 0.3922 & 0.1672 & 0.8551 & 0.5599 & 0.7116 & 0.6735 & 0.5012 & 0.6342 & 0.5352 \\
\textit{CLaMP 3$_{as}$} & 0.9097 & 0.3888 & 0.4935 & 0.8379 & 0.5944 & 0.7413 & 0.6445 & 0.8601 & 0.6780 & 0.8491 \\
\textit{CLaMP 3$_{sa}$} & 0.9084 & 0.3863 & 0.2533 & 0.8448 & \textbf{0.6031} & 0.6949 & 0.6338 & 0.8647 & 0.7061 & 0.8419 \\
\textit{CLaMP 3$_{sa}^{c2}$} & 0.9092 & 0.3924 & 0.2545 & 0.8551 & 0.5477 & 0.6876 & 0.6147 & 0.8574 & 0.6710 & 0.8007 \\
\textit{CLaMP 3$_{assa}$} & 0.9098 & 0.3935 & 0.1498 & \textbf{0.8793} & 0.5921 & 0.7327 & 0.6411 & 0.8742 & 0.6842 & 0.8555 \\
\textit{CLaMP 3$_{saas}$} & \textbf{0.9109} & \textbf{0.3941} & 0.5377 & 0.8655 & 0.5907 & 0.7004 & 0.6377 & 0.8689 & 0.7053 & 0.8441 \\
\textit{CLaMP 3$_{saas}^{c2}$} & 0.9095 & 0.3938 & 0.3907 & 0.8138 & 0.5368 & 0.6589 & 0.6562 & 0.8732 & 0.6798 & 0.8470 \\
\bottomrule
\end{tabular}
\label{tab:marble_cls}
\end{table*}

\begin{table*}[t]
\centering
\renewcommand{\arraystretch}{1}
\setlength{\tabcolsep}{10pt}
\fontsize{9.7}{10}\selectfont
\caption{Audio classification performance on the MTG-Jamendo dataset (55,000+ tracks) was evaluated across four tasks: instrument classification (41 tags), mood/theme classification (59 tags), genre classification (95 tags), and top-50 multi-label classification.}
\begin{tabular}{l c c c c c c c c}
\toprule
\multirow{2}{*}{\textbf{\vspace{-2mm}Model}} & 
\multicolumn{2}{c}{\textbf{Instrument}} & \multicolumn{2}{c}{\textbf{Mood/Theme}} & \multicolumn{2}{c}{\textbf{Genre}} & \multicolumn{2}{c}{\textbf{Top50}} \\
\cmidrule(lr){2-3} \cmidrule(lr){4-5} \cmidrule(lr){6-7} \cmidrule(lr){8-9}
& \textit{ROC} & \textit{AP} & \textit{ROC} & \textit{AP} & \textit{ROC} & \textit{AP} & \textit{ROC} & \textit{AP} \\
\midrule
\textit{MERT$_{mean}$} & 0.7421 & 0.1764 & 0.7598 & 0.1383 & 0.8672 & 0.1818 & 0.8280 & 0.2837 \\
\textit{CLAP} & 0.7480 & 0.1812 & 0.7601 & 0.1323 & 0.8544 & 0.1716 & 0.8197 & 0.2773 \\
\textit{TTMR++} & 0.7806 & 0.2111 & 0.7705 & 0.1477 & 0.8742 & 0.2030 & \textbf{0.8340} & 0.3049 \\
\textit{CLaMP 3$_{as}$} & 0.7895 & 0.2254 & 0.7814 & 0.1476 & 0.8750 & \textbf{0.2114} & 0.8321 & 0.3068 \\
\textit{CLaMP 3$_{sa}$} & 0.7780 & 0.2112 & 0.7823 & 0.1533 & 0.8713 & 0.2008 & 0.8276 & 0.3011 \\
\textit{CLaMP 3$_{sa}^{c2}$} & 0.7832 & 0.2168 & 0.7796 & 0.1475 & 0.8679 & 0.2046 & 0.8220 & 0.2964 \\
\textit{CLaMP 3$_{assa}$} & \textbf{0.7911} & \textbf{0.2269} & 0.7828 & 0.1486 & \textbf{0.8763} & 0.2109 & 0.8290 & 0.3041 \\
\textit{CLaMP 3$_{saas}$} & 0.7872 & 0.2208 & \textbf{0.7835} & \textbf{0.1547} & 0.8703 & 0.2076 & 0.8242 & 0.3021 \\
\textit{CLaMP 3$_{saas}^{c2}$} & 0.7803 & 0.2145 & 0.7825 & 0.1522 & 0.8734 & 0.2092 & 0.8296 & \textbf{0.3074} \\
\bottomrule
\end{tabular}
\label{tab:mtg_cls}
\end{table*}

\clearpage

\section{Music Classification}
\label{sec:classification}
This section evaluates CLaMP 3 variants and baselines via linear probing, assessing their ability to classify musical attributes in symbolic and audio music, as well as musical modalities and text annotations in WikiMT-X.

\subsection{Symbolic Music Classification}
Table~\ref{tab:symbolic_cls} presents symbolic music classification results for ABC notation and MIDI across three benchmarks:

\textbf{WikiMT} \cite{DBLP:conf/ismir/WuY0S23} consists of 1,010 lead sheets in ABC notation sourced from Wikifonia\footnote{\url{http://www.synthzone.com/files/Wikifonia/Wikifonia.zip}}, labeled into 8 genre categories based on corresponding Wikipedia entries.

\textbf{VGMIDI} \cite{DBLP:conf/ismir/FerreiraW19} contains 204 MIDI transcriptions of video game soundtracks, annotated with 4 emotion labels derived from valence and arousal levels.

\textbf{Pianist8} \cite{chou2021midibert} includes 411 piano performances, transcribed from audio to performance MIDI, and labeled with their respective composers across eight categories.

To enable evaluation in both formats, all datasets were converted between ABC and MIDI.

Despite improved text alignment, CLaMP 3 does not surpass CLaMP 2 in sheet music classification. This is likely because CLaMP 3 was trained on only half as much symbolic data. While stronger textual supervision benefits retrieval, it does not fully offset the reduced symbolic training for classification. However, CLaMP 3 still outperforms M3—the symbolic music encoder it was initialized from—on most benchmarks, suggesting that contrastive text supervision enhances the semantic salience of extracted features.

These results indicate that retrieval and classification improvements are relatively independent. In text-to-music retrieval (Table~\ref{tab:model_comparison}, Table~\ref{tab:multilingual_comparison}), CLaMP 3—especially CLaMP 3\(_{\textnormal{sa}}^{\textnormal{c2}}\)—significantly outperforms CLaMP 2, yet this advantage does not extend to classification. A possible explanation is that retrieval requires rich representations and effective interaction between text and music encoders, while classification depends solely on an encoder’s ability to extract features relevant to predefined labels. Thus, while higher-quality text annotations enhance retrieval, they do not necessarily improve symbolic music classification.

\subsection{Audio Music Classification}
To evaluate the audio classification performance of CLaMP 3 variants and baselines, we conduct linear probing on MARBLE \cite{DBLP:conf/nips/YuanMLZCYZLHTDW23} and MTG-Jamendo \cite{bogdanov2019mtg}.

MARBLE is a comprehensive benchmark collection for music representation evaluation. We assess models on 8 tasks covering different aspects of audio understanding. MTG-Jamendo is a large-scale benchmark with over 55,000 music tracks annotated for multiple classification tasks. It focuses on high-level musical attributes, making it well-suited for evaluating a model’s ability to capture semantic meaning in music.

We also assess the self-supervised model MERT, CLaMP 3’s audio feature extractor, averaging embeddings to one per 5-second clip across layers and time steps.

Table~\ref{tab:marble_cls} shows the strengths of contrastive and self-supervised models on the MARBLE benchmark. CLaMP 3 variants excel in high-level tasks, like genre classification (GTZAN) and tagging (MTT), where capturing abstract musical meaning is crucial. MERT, however, performs better in low-level tasks such as key detection (GS) and pitch classification (Nsynth), where fine spectral detail is more important. Contrastive models generally struggle with short-duration audio (e.g., 4-second clips in Nsynth) because their focus on aligning longer segments with text limits their ability to capture acoustic details. These results suggest contrastive learning is better for semantic tasks, while self-supervised models are more effective for low-level acoustic analysis.

Table~\ref{tab:mtg_cls} shows that contrastive models, particularly CLaMP 3 variants, consistently outperform MERT across all MTG-Jamendo tasks. Notably, CLaMP 3 models achieve the highest scores in most tasks, demonstrating how diverse and high-quality text annotations help contrastive models learn and capture complex musical semantics.

In summary, contrastive models perform well in high-level classification tasks but struggle with short clips and fine-grained acoustic details. Their effectiveness heavily depends on the text annotations used during training. For instance, CLAP achieves strong results in instrument classification (Nsynth) because its training data is dominated by instrument and genre descriptions. However, it performs poorly in key detection (GS), where such annotations offer little relevant information.

\clearpage

\twocolumn[{
\noindent
\begin{minipage}{\textwidth}
    \centering
    \renewcommand{\arraystretch}{1}
    \setlength{\tabcolsep}{10pt}
    \fontsize{9.7}{10}\selectfont
    \captionof{table}{Classification performance on WikiMT-X (1,000 entries, 8 genres) across different musical modalities and text annotations.}
    \label{tab:wikimtx_cls}
    \begin{tabular}{l c c c c c c c}
    \toprule
    \textbf{Model} & \textbf{ABC} & \textbf{MIDI} & \textbf{Audio} & \textbf{Background} & \textbf{Analysis} & \textbf{Description} & \textbf{Scene} \\
    \midrule
    \textbf{Accuracy} & & & & & & & \\
    \textit{CLaMP} & 0.7000 & - & - & 0.8050 & 0.7900 & 0.6900 & 0.6250 \\
    \textit{CLaMP 2} & 0.6800 & 0.6350 & - & 0.7900 & 0.8150 & 0.7250 & 0.6150 \\
    \textit{CLAP} & - & - & 0.6450 & 0.6950 & 0.6800 & 0.6500 & 0.5550 \\
    \textit{TTMR++} & - & - & 0.7150 & 0.7400 & 0.7600 & 0.6700 & 0.5950 \\
    \textit{CLaMP 3$_{as}$} & 0.6850 & 0.6100 & 0.7050 & 0.8200 & 0.8350 & \textbf{0.7800} & 0.6550 \\
    \textit{CLaMP 3$_{sa}$} & 0.7000 & 0.6650 & 0.6850 & 0.8000 & 0.8600 & 0.7700 & 0.6500 \\
    \textit{CLaMP 3$_{sa}^{c2}$} & 0.6850 & 0.6350 & 0.6850 & 0.7850 & 0.8550 & 0.7750 & 0.6500 \\
    \textit{CLaMP 3$_{assa}$} & 0.7000 & 0.6300 & \textbf{0.7200} & \textbf{0.8650} & \textbf{0.8650} & 0.7700 & \textbf{0.6850} \\
    \textit{CLaMP 3$_{saas}$} & \textbf{0.7150} & \textbf{0.6800} & 0.7050 & 0.8400 & 0.8550 & \textbf{0.7800} & 0.6650 \\
    \textit{CLaMP 3$_{saas}^{c2}$} & 0.6750 & 0.6300 & 0.6850 & 0.8300 & 0.8500 & 0.7700 & \textbf{0.6850} \\
    \midrule
    \midrule
    \textbf{F1-macro} & & & & & & & \\
    \textit{CLaMP} & 0.5252 & - & - & 0.6835 & 0.6486 & 0.6079 & 0.4447 \\
    \textit{CLaMP 2} & 0.5287 & 0.3784 & - & 0.6617 & 0.6832 & 0.6333 & 0.3710 \\
    \textit{CLAP} & - & - & 0.3943 & 0.5913 & 0.5491 & 0.4921 & 0.3100 \\
    \textit{TTMR++} & - & - & 0.4714 & 0.6914 & 0.6694 & 0.6254 & 0.4246 \\
    \textit{CLaMP 3$_{as}$} & 0.5431 & 0.4005 & 0.4755 & 0.7424 & 0.7933 & \textbf{0.7639} & 0.4780 \\
    \textit{CLaMP 3$_{sa}$} & 0.5345 & 0.5108 & 0.4881 & 0.7917 & 0.8199 & 0.7372 & 0.4527 \\
    \textit{CLaMP 3$_{sa}^{c2}$} & 0.5428 & 0.4171 & 0.4589 & 0.6626 & 0.7439 & 0.7318 & 0.4260 \\
    \textit{CLaMP 3$_{assa}$} & 0.5499 & 0.3976 & \textbf{0.5130} & \textbf{0.8486} & \textbf{0.8277} & 0.6878 & \textbf{0.5207} \\
    \textit{CLaMP 3$_{saas}$} & \textbf{0.5720} & \textbf{0.4967} & 0.4995 & 0.8123 & 0.8225 & 0.7484 & 0.4742 \\
    \textit{CLaMP 3$_{saas}^{c2}$} & 0.5182 & 0.4313 & 0.4432 & 0.7811 & 0.8054 & 0.7082 & 0.4999 \\
    \bottomrule
    \end{tabular}
    \vspace{2em}
\end{minipage}
}]

\subsection{Classification on WikiMT-X}
Table~\ref{tab:wikimtx_cls} presents classification results across different musical modalities (ABC, MIDI, Audio) and text annotations (\textit{Background}, \textit{Analysis}, \textit{Description}, \textit{Scene}) on WikiMT-X.

Compared to the WikiMT results in Table~\ref{tab:symbolic_cls}, all models show substantial gains in genre classification accuracy and F1-macro for ABC and MIDI. This confirms that reannotating genre labels significantly reduced label noise, leading to more reliable classification. The improvements suggest that earlier inconsistencies in genre annotations were a major limiting factor in classification performance. The reorganized label taxonomy and refined annotations in WikiMT-X provide a more structured and consistent genre framework, making it a more reliable benchmark for music classification.

Across different musical modalities, the best-performing models for ABC, MIDI, and Audio achieve comparable classification results. This suggests that genre-related features are well-preserved regardless of musical representation. Fig.~\ref{figure:t-sne,a} further supports this observation, showing clear genre boundaries across all modalities, indicating CLaMP 3 models can effectively extract genre information from both representations, reinforcing the idea that genre characteristics are consistently encoded in musical data.

A clear distinction emerges between text and music classification: models perform significantly better on text annotations (\textit{Background}, \textit{Analysis}, \textit{Description}) than on music data. This is likely because text often contains explicit genre-related cues, making classification more direct. For example, descriptions like “syncopated piano chords and walking bass” strongly suggest jazz. In contrast, classifying music requires models to infer genre from intricate relationships between harmony, rhythm, and timbre. However, \textit{Scene} classification behaves differently from other text-based categories—it describes environmental settings rather than musical attributes, making its classification challenge more similar to music than text.

Models trained solely on audio-text alignment (i.e., CLAP, TTMR++) perform worse in text classification, likely due to the limited diversity of annotations in large-scale audio-text datasets, which often list only instruments and genres. In contrast, symbolic-text datasets provide richer semantics, including background context and musicological analysis. CLaMP 3\(_{\textnormal{as}}\) is an exception—though its text encoder was fully updated during audio alignment, it achieves much stronger text classification than models like CLAP and TTMR++. This is likely due to M4-RAG’s well-curated and diverse annotations, which offer a broader and more expressive linguistic representation of musical content.

\clearpage

\twocolumn[{
\noindent
\begin{minipage}{\textwidth}
    \centering
    \renewcommand{\arraystretch}{1.25}
    \setlength{\tabcolsep}{10pt}
    \fontsize{9.7}{9}\selectfont
    \captionof{table}{Results for English text-to-music retrieval on MusicCaps, reflecting data leakage in baseline models. Evaluations are conducted on both the full set and the AudioSet evaluation set. R/O denotes the use of rewritten or original captions, while F/C indicates retrieval using full tracks or clips.}
    \label{tab:musiccaps_leakage}
    \begin{tabular}{l c c c c c c c c}
    \toprule
    \multirow{2}{*}{\vspace{-2mm}\textbf{Model}} & \multicolumn{4}{c}{\textbf{Full Set (5,521 pairs)}} & \multicolumn{4}{c}{\textbf{Eval Set (2,858 pairs)}} \\
    \cmidrule(lr){2-5} \cmidrule(lr){6-9}
    & \textit{RF} & \textit{RC} & \textit{OF} & \textit{OC} & \textit{RF} & \textit{RC} & \textit{OF} & \textit{OC} \\
    \midrule
    \textit{CLAP} & 0.0536 & 0.0743 & 0.0640 & 0.0894 & 0.0657 & 0.0886 & 0.0774 & 0.1113 \\
    \textit{TTMR++} & \textbf{0.1410} & \textbf{0.2315} & \textbf{0.1757} & \textbf{0.3155} & \textbf{0.1248} & \textbf{0.1341} & \textbf{0.1219} & \textbf{0.1382} \\
    \textit{CLaMP 3$_{as}$} & 0.0874 & 0.0642 & 0.0696 & 0.0536 & 0.1119 & 0.0830 & 0.0917 & 0.0699 \\
    \textit{CLaMP 3$_{sa}$} & 0.0741 & 0.0591 & 0.0530 & 0.0431 & 0.0934 & 0.0735 & 0.0661 & 0.0572 \\
    \textit{CLaMP 3$_{sa}^{c2}$} & 0.0729 & 0.0609 & 0.0619 & 0.0504 & 0.0961 & 0.0832 & 0.0822 & 0.0651 \\
    \textit{CLaMP 3$_{assa}$} & 0.0830 & 0.0592 & 0.0743 & 0.0530 & 0.1045 & 0.0784 & 0.0897 & 0.0723 \\
    \textit{CLaMP 3$_{saas}$} & 0.0890 & 0.0705 & 0.0652 & 0.0523 & 0.1177 & 0.0889 & 0.0890 & 0.0682 \\
    \textit{CLaMP 3$_{saas}^{c2}$} & 0.0973 & 0.0737 & 0.0762 & 0.0550 & 0.1180 & 0.0933 & 0.0961 & 0.0710 \\
    \bottomrule
    \end{tabular}
    \vspace{2em}
\end{minipage}
}]

\section{Data Leakage of MusicCaps}
\label{sec:musiccaps}
MusicCaps, a widely used text-to-music retrieval benchmark, includes 5,521 music-text pairs with 10-second audio clips. As a subset of AudioSet, many models are trained on overlapping data, raising concerns about reliability, as they may memorize seen examples rather than learning true retrieval patterns.

Table~\ref{tab:musiccaps_leakage} shows text-to-music retrieval results on MusicCaps, examining data leakage in baseline models. We evaluate performance on the full dataset (Full Set) and the AudioSet evaluation subset (Eval Set), while also assessing the effects of caption rewording (Original vs. Rewritten) and audio length (Clip vs. Full Track).

Leakage varies across models: TTMR++ is the most affected, having been trained on MusicCaps pairs from the training set of AudioSet, exposing it to half the benchmark; CLAP, trained on the full AudioSet, has seen all MusicCaps audio; in contrast, CLaMP 3 has minimal exposure, with only 150 audio recordings appearing in M4-RAG.

To mitigate leakage effects, we introduce rewritten captions generated using Qwen, ensuring semantic consistency while incorporating structured aspect lists—detailed annotations of key musical attributes such as instrumentation, mood, and rhythm. Additionally, we conduct retrieval on both 10-second clips and full-length tracks, forming four evaluation settings:

\begin{itemize}
    \item \textbf{RF:} Rewritten captions with full tracks.
    \item \textbf{RC:} Rewritten captions with clips.
    \item \textbf{OF:} Original captions with full tracks.
    \item \textbf{OC:} Original captions with clips.
\end{itemize}

Table~\ref{tab:musiccaps_leakage} reveals clear data leakage. TTMR++ is the only model that performs worse on the evaluation set than on the full benchmark, despite the evaluation set containing fewer retrieval candidates, which should naturally lead to higher MRR scores. This suggests severe overfitting to seen MusicCaps training data. Additionally, both TTMR++ and CLAP show performance drops with rewritten captions and full-length tracks. For TTMR++, this suggests that these modifications help reduce leakage effects, though not entirely. For CLAP, the decline is likely due to rewritten captions incorporating more detailed semantic information from aspect lists, which may shift retrieval behavior.

In contrast, all CLaMP 3 variants show improved performance with rewritten captions, likely due to M4-RAG’s use of Qwen, making them more attuned to its text patterns. They also gain an advantage in full-track retrieval. While baseline models rely on 10-second clips and average embeddings across segments, CLaMP 3 processes up to 640 seconds of audio, enabling it to capture relationships across an entire track. In contrast, baselines extract semantics from isolated clips, restricting their ability to utilize long-form audio context effectively.

These results raise broader concerns about benchmark reliability in text-to-music retrieval. Other benchmarks also face leakage risks—SDD, for instance, comes from MTG-Jamendo, which was included in CLAP’s training data. In contrast, WikiMT-X, manually curated for this study, mitigates leakage by sourcing audio from the web rather than existing datasets. However, since this audio remains publicly accessible, large-scale models may still have exposure. To further reduce leakage, future benchmarks should prioritize private or newly recorded datasets for unbiased evaluation.
\end{document}